\documentclass[aps,prd,amsmath,amsfonts,amssymb,eqsecnum,nofootinbib,floatfix,secnumarabic,preprintnumbers,superscriptaddress,nobalancelastpage,onecolumn,notitlepage,biblatex,12pts]{article}
\pdfoutput=1 

\usepackage[T1]{fontenc} 

\usepackage[]{caption}
\usepackage{makeidx}
\usepackage{multirow}
\usepackage{multicol}
\usepackage[dvipsnames,svgnames,table]{xcolor}
\usepackage{graphics}
\usepackage{epstopdf}
\usepackage{ulem}
\usepackage{hyperref}
\usepackage{amsmath}
\usepackage[compat=1.0.0]{tikz-feynman}
\usepackage{amssymb}
\usepackage{float}
\usepackage{fancyhdr} 
\usepackage{enumitem}
\usepackage{tcolorbox}
\usepackage{lipsum}
\usepackage{bold-extra}
\usepackage{wrapfig}
\usepackage{caption}
\usepackage[numbers]{natbib}
\allowdisplaybreaks
\usepackage{upgreek}

\usepackage{bm}					
\usepackage{booktabs} 

\newcommand{\be}{\begin{eqnarray}}
\newcommand{\ee}{\end{eqnarray}}
\newcommand{\bea}{\begin{eqnarray}}
\newcommand{\eea}{\end{eqnarray}}

\newcommand{\ba}{\begin{array}}
\newcommand{\ea}{\end{array}}
\newcommand{\bd}{\begin{displaymath}}
\newcommand{\ed}{\end{displaymath}}
\newcommand{\beq}{\begin{equation}}
\newcommand{\eeq}{\end{equation}}

\def\q2 {q^2}

\def\bt{\begin{table}}
\def\et{\end{table}}

\definecolor{shilamagenta}{rgb}{0.8, 0.0, 0.8}

\definecolor{shilagreen}{rgb}{0.0, 0.5, 0.0}
\definecolor{shilacyan}{rgb}{0.0, 0.58, 0.71}

\definecolor{midnightblue}{rgb}{0.1, 0.1, 0.44}

\usepackage{hyperref}

\hypersetup{citecolor=blue,        
    linkcolor=red,  
    filecolor=magenta,      
    urlcolor=shilacyan  
}

\usepackage{cleveref}
\usepackage{slashed}

\begin{document}

\title{\textcolor{black}{\textbf{Gravitational Waves from First-Order Phase Transitions Assisted by Temperature-Enhanced Scatterings}}}

\author{Arnab~Chaudhuri
\footnote{{\bf e-mail}: \href{arnab.chaudhuri@nao.ac.jp}{arnab.chaudhuri@nao.ac.jp} }\\
\small{Division of Science, National Astronomical Observatory of Japan, Mitaka, Tokyo 181-8588, Japan.} 
}

\date{\today}
\maketitle

\begin{abstract}
Scatterings whose cross sections increase as the cosmic temperature decreases, known as temperature - enhanced scatterings, can have a significant impact on the thermal effective potential of scalar fields responsible for driving cosmological first-order phase transitions. We show that such effects naturally manifest as finite-temperature self-energy corrections to the scalar mass term, leading to an additional contribution of the form \(c\,T^{p}\phi^{2}\) in the effective potential. In this work, we systematically investigate how these loop-induced, temperature-dependent corrections affect key phase transition parameters, including the nucleation temperature, latent heat release, and inverse duration parameter. These modifications influence both the strength and duration of the phase transition, which in turn determine the properties of the resulting stochastic gravitational-wave (GW) background. Employing semi-analytic computational methods, we evaluate the GW spectra generated under these conditions and compare our predictions with the projected sensitivities of forthcoming detectors such as LISA, DECIGO, and BBO. Our analysis demonstrates that finite-temperature scattering effects of this kind can substantially strengthen first-order transitions and produce GW signals that lie within the reach of future observational facilities. The results establish a concrete thermal-field-theoretic origin for temperature-dependent modifications of the scalar potential and emphasize their importance in shaping early-Universe cosmological signatures.
\\
\\
\textbf{Keywords:} Gravitational Waves, Cosmological Phase Transitions, Early Universe.
\end{abstract}

\section{Introduction}
\label{sec:intro}

The origin of the baryon asymmetry of the Universe (BAU) remains one of the most compelling open questions in modern cosmology and particle physics~\cite{Sakharov:1967dj, Dine:2003ax, Riotto:1999yt, Buchmuller:2005eh, Morrissey:2012db, Canetti:2012zc}.
Numerous mechanisms have been proposed, including electroweak baryogenesis~\cite{Kuzmin:1985mm, Cohen:1993nk, Cline:2006ts, Morrissey:2012db, Patel:2011th, Konstandin:2013caa, Carena:2008rt}, leptogenesis~\cite{Fukugita:1986hr, Davidson:2008bu, Buchmuller:2005eh, Blanchet:2012bk, Hambye:2004jf, Hambye:2012fh, Dev:2017trv}, Affleck–Dine baryogenesis~\cite{Affleck:1984fy, Dine:1995kz, Kusenko:1997si}, and mechanisms involving Q-balls~\cite{Kusenko:1997zq, Enqvist:1998xd, Kasuya:1999wu}, but the precise dynamics that produced the observed asymmetry remain elusive.

A recent proposal~\cite{Flores:2025fvd} has highlighted an alternative possibility: baryogenesis through \emph{critical and super-critical scatterings}, where the cross sections for interactions between non-relativistic particles and the thermal bath increase as the Universe cools.
Such temperature-enhanced scatterings can sustain or even amplify asymmetry generation at late times, circumventing the limitations faced by conventional scenarios.

In parallel, first-order phase transitions (FOPTs) in the early Universe have attracted significant interest as potential sources of a stochastic gravitational wave (GW) background~\cite{Chaudhuri:2022sis,Witten:1984rs, Hogan:1986qda, Kosowsky:1991ua, Kosowsky:1992rz, Kamionkowski:1993fg, Caprini:2015zlo, Caprini:2019egz, Mazumdar:2018dfl, Caprini:2009fx, Hindmarsh:2017gnf,Athron:2023xlk,Ghosh:2022fzp,Bittar:2025lcr}.
A strongly first-order phase transition proceeds through bubble nucleation, expansion, and collisions, together with plasma processes such as sound waves and turbulence~\cite{Huber:2008hg, Hindmarsh:2013xza, Hindmarsh:2015qta, Weir:2017wfa, Cutting:2018tjt, Jinno:2016vai, Jinno:2017ixd, Ellis:2018mja, Caprini:2019egz, Cutting:2019zws}.
These mechanisms can generate GW signals within the reach of planned experiments such as LISA~\cite{LISA:2017pwj, Caprini:2019egz, Schmitz:2020syl}, DECIGO~\cite{Seto:2001qf, Kawamura:2020pcg}, BBO~\cite{Corbin:2005ny}, ET~\cite{Punturo:2010zz}, and CE~\cite{Evans:2021gyd}.

It is therefore natural to ask whether the same temperature-enhanced scatterings that can drive baryogenesis might also affect the dynamics of a FOPT, potentially strengthening the transition and enhancing its GW signal.
Such scatterings modify the thermal effective potential of the scalar field driving the phase transition, thereby altering the critical and nucleation temperatures, the latent heat released, and the duration of the transition.
These parameters directly control the amplitude and spectral shape of the resulting GW background~\cite{Caprini:2015zlo, Caprini:2019egz, Mazumdar:2018dfl, Grojean:2006bp, Espinosa:2010hh, Patel:2011th, Kakizaki:2015wua, Huang:2017rzf, Beniwal:2017eik, Alves:2018jsw, Athron:2019teq, Hashino:2018wee, Chiang:2018gsn, Balazs:2016tbi, Chala:2016ykx, Croon:2020cgk, Ramsey-Musolf:2019lsf, Fairbairn:2019xog, Ellis:2019oqb,Chattopadhyay:2022fwa}.

From the standpoint of finite-temperature quantum field theory, temperature-enhanced scatterings correspond to loop-level corrections to the scalar two-point function.  The resulting finite-$T$ self-energy $\Pi_\phi(T)$ modifies the thermal mass and hence the curvature of the effective potential near the origin.  In the local limit of the one-particle-irreducible effective action, this effect appears as an additional quadratic term in the potential, $\Delta V(\phi,T)\propto \Pi_\phi(T)\,\phi^2$.  Depending on the mediator spectrum and the thermal scaling of the underlying cross sections, the self-energy can behave as $\Pi_\phi(T)\!\propto\! T^{p}$ with either positive or negative exponent $p$, thereby motivating the phenomenological parameterization adopted below.  This correction is conceptually distinct from the collision terms that enter Boltzmann equations for wall friction or transport---here we focus exclusively on its static contribution to the effective potential.

A temperature-dependent quadratic correction of this form arises naturally in several concrete particle-physics models.  
In particular, singlet-extended electroweak scenarios (xSM) generate both (i) a tree-level barrier capable of producing a first-order electroweak phase transition, and (ii) loop-induced thermal self-energy corrections that modify the Higgs curvature as $T$ evolves.  
The combination of Higgs–singlet mixing and thermal dressing of the singlet propagator yields precisely a correction of the structure $\Delta V(\phi,T)=c\,T^{p}\phi^{2}$, with the exponent $p$ determined by the dominant microscopic scattering processes.  
A representative realization of this mechanism is developed explicitly in Sec.~\ref{sec:UVexampledetail}, where the mapping between the microscopic Lagrangian and the effective parameters $(c,p)$ is derived from the corresponding thermal loop diagrams.

In this work, we explore this connection in a general phenomenological framework.
Rather than committing to a specific UV-complete particle physics model, we parametrize the impact of temperature-enhanced scatterings on the thermal effective potential by introducing an additional term,
\begin{equation} \label{eq:delV}
\Delta V(\phi, T) = c\, T^{p} \phi^2,
\end{equation}
where $p<0$ captures the growing impact of scatterings as the temperature decreases.
The full finite-temperature potential used in this study thus reads
\begin{equation}
V_{\text{eff}}(\phi,T)=V_{0}(\phi)+V_{\text{CW}}(\phi)+V_{T}(\phi,T)+c\,T^{p}\phi^{2},
\end{equation}
where $V_{0}(\phi)$ denotes the tree-level potential of the model, $V_{\text{CW}}$ the Coleman--Weinberg one-loop term, and $V_{T}$ the usual thermal contribution from particles in equilibrium.  Explicit expressions are provided in Sec.~\ref{sec:framework}.
We analyze how this term shifts key phase transition parameters, namely the nucleation temperature $T_n$, the strength parameter $\alpha$, and the inverse duration parameter $\beta$~\cite{Caprini:2015zlo, Caprini:2019egz, Grojean:2006bp}.

We then compute the resulting GW spectrum using standard semi-analytic techniques~\cite{Huber:2008hg, Caprini:2015zlo, Caprini:2019egz, Hindmarsh:2017gnf, Ellis:2018mja}.
By scanning the parameter space $(c,p)$, we identify regions where the GW signal may be within reach of planned observatories such as LISA, DECIGO, and BBO.

Our results demonstrate that temperature-enhanced scatterings can significantly amplify a FOPT by increasing its strength parameter $\alpha$ and reducing its inverse duration parameter $\beta$, thereby generating an observable GW signal.
This provides an intriguing complementarity: the same mechanism that could produce the BAU may also lead to testable predictions for GW experiments.
Moreover, by grounding the $T$-dependent correction in explicit finite-temperature self-energy effects, our analysis connects microphysical scattering processes to macroscopic cosmological observables in a self-consistent manner.
This connection highlights the power of GW cosmology as a probe of physics beyond the Standard Model and motivates further exploration of explicit particle models that realize this scenario~\cite{Croon:2020cgk, Athron:2019teq, Alves:2018jsw, Beniwal:2017eik, Hashino:2018wee, Chiang:2018gsn, Balazs:2016tbi, Chala:2016ykx, Ellis:2019oqb, Fairbairn:2019xog, Ramsey-Musolf:2019lsf, Srivastava:2025oer, Chaudhuri:2024vrd, Roy:2022gop, Chatterjee:2022pxf, Borah:2023zsb,Ghosh:2025cxp,Chaudhuri:2025ylu,Chaudhuri:2025cjp,Chaudhuri:2021agl,Chaudhuri:2021rwt,Chaudhuri:2021ibc}.

This paper is organized as follows.
In Section~\ref{sec:framework}, we present the theoretical framework and parameterization of temperature-enhanced scatterings and describe their effect on the thermal effective potential.
Section~\ref{sec:gw} outlines the methodology used to compute the associated stochastic GW spectrum.
In Section~\ref{sec:parameter_scan}, we examine the parameter space and evaluate the GW signal for representative benchmark points.
Our main results are presented and discussed in Section~\ref{sec:results}, followed by our summary and outlook in Section~\ref{sec:discussion_conclusion}.


\section{Framework and methodology}
\label{sec:framework}

In this section, we describe the theoretical setup used to explore the impact of temperature-enhanced scatterings on cosmological first-order phase transitions and the resulting GW signals.
Rather than constructing a specific ultraviolet-complete particle physics model, we adopt a phenomenological approach that captures the essential physics and allows us to scan systematically over the relevant parameter space.

\subsection{Temperature-enhanced scatterings}
\label{subsec:scatterings}

Recently, it has been proposed that the baryon asymmetry of the Universe may originate from scatterings whose cross sections increase as the temperature of the Universe decreases~\cite{Flores:2025fvd}.
Such scatterings can be described by a temperature-dependent cross section of the form
\begin{equation}
\sigma(T) \propto T^{-n},
\end{equation}
with $n>0$, so that the cross section becomes larger at lower temperatures.
This scaling may arise, for example, in models with light mediators, composite states, or nonperturbative effects that become important as the thermal bath cools.

The key insight of~\cite{Flores:2025fvd} is that these temperature-enhanced scatterings can sustain or even increase baryogenesis efficiency at late times, circumventing the usual washout and Boltzmann suppression that affect standard scenarios.
We explore whether the same scatterings, through their impact on the thermal effective potential, can also modify the dynamics of a first-order phase transition, potentially leading to observable GW signals.

It is important to emphasize that a temperature–dependent interaction rate enters the effective potential only through equilibrium self-energy corrections, i.e. the real part of the thermal two-point function. Collision terms appearing in Boltzmann transport equations describe non-equilibrium frictional effects and do not alter the free energy of the system. In our treatment the latter are handled separately through the bubble-wall velocity, while the equilibrium self-energy contribution modifies $V_{\mathrm{eff}}$ via a local $\phi^{2}$ operator.

\subsection{Parametrization of the thermal effective potential}
\label{subsec:potential}

To study the impact of temperature-enhanced scatterings on the scalar field $\phi$ responsible for the phase transition, we parametrize the thermal effective potential as
\begin{equation} \label{eq:Veff}
V_{\mathrm{eff}}(\phi,T) = V_0(\phi) + \Delta V(\phi,T),
\end{equation}
where $V_0(\phi)$ is the zero-temperature potential of the model and $\Delta V(\phi,T)$ encodes finite-temperature corrections.


The standard one-loop corrections are included via the Coleman–Weinberg and thermal integrals,
\begin{equation}
V_{\rm CW}(\phi)=\sum_i \frac{n_i}{64\pi^2} m_i^4(\phi)
\Big[\ln\frac{m_i^2(\phi)}{\mu^2}-C_i\Big],
\quad
V_T(\phi,T)=\sum_i \frac{n_i T^4}{2\pi^2} J_{B/F}\!\left(\frac{m_i^2(\phi)}{T^2}\right),
\end{equation}
where $n_i$ counts degrees of freedom, $m_i(\phi)$ are field-dependent masses, and $J_{B/F}$ are the standard bosonic/fermionic thermal functions~\cite{Quiros:1999jp}.

All thermal functions $J_{B,F}(m^2/T^2)$ are evaluated using the full integral expressions,
without invoking any high- or low-temperature series expansions. We follow the numerical
prescription of Ref.~\cite{Bernon:2017jgv}, which directly integrates the exact
thermal potentials and therefore remains valid in the domain $m^2/T^2=\mathcal{O}(1)$ where
asymptotic expansions fail. This guarantees that the scattering-induced term $cT^{p}\phi^{2}$
supplements, but does not double count, the equilibrium thermal contributions.

We emphasize that different sectors of the theory are treated in different thermal regimes.
The cubic term and tree-level barrier responsible for the first-order phase transition arise
from light degrees of freedom (such as gauge bosons and scalars with field-dependent masses
$m_i(\phi)\lesssim T$ near the transition), for which the standard high-temperature expansion
of the thermal functions is valid.
In contrast, the mediator field $S$ responsible for the temperature-enhanced scattering
contribution is not assumed to be light.
Its effect is incorporated solely through its static, thermally dressed propagator entering
the scalar self-energy, without invoking a high-temperature expansion for $S$ itself.
As a result, the assumptions $m_i(\phi)\ll T$ for barrier-generating fields and
$m_S \gtrsim T$ for the mediator are not in conflict.

\subsubsection{Thermal self-energy origin of the \texorpdfstring{$cT^{p}\phi^{2}$}{cTp phi2} term}

From thermal field theory, the leading finite-temperature correction to a scalar field arises from the one-loop self-energy
\begin{equation}
\Pi_\phi(T) = \mathrm{Re}\,\Sigma_\phi(p_0=0,\mathbf p\!\to\!0;T),
\end{equation}
which modifies the effective potential as
\begin{equation}
\Delta V(\phi,T)\simeq \tfrac12 \Pi_\phi(T)\,\phi^2.
\end{equation}
For heavy mediators one obtains $\Pi_\phi(T)\propto +T^2$—the usual positive thermal mass.
For light $t$-channel mediators, however, the typical momentum transfer scales as $|t|\!\sim\!T^2$ and the loop amplitude behaves as $\sim1/(T^2+m_S^2)^2$, giving $\Pi_\phi(T)\propto T^{p}$ with negative $p$ in the regime $T\gg m_S$.

In this local, static limit the contribution is a genuine equilibrium correction to the free energy, and does not reproduce any transport/friction effect that would otherwise enter the Boltzmann equations for bubble-wall propagation. Hence no double counting occurs when the mediator is not already part of the thermal particle content entering $V_T(\phi,T)$.

The additional term $cT^{p}\phi^{2}$ represents the equilibrium self-energy contribution of a specific microscopic degree of freedom that is not part of the thermal sum defining $V_{T}(\phi,T)$. Since $V_{T}$ includes only fields with thermal occupation in the plasma, adding $cT^{p}\phi^{2}$ does not duplicate any existing contribution. Non-equilibrium effects such as friction arise only in transport equations and are treated independently in Sec.~\ref{subsec:setup}, further ensuring that equilibrium and non-equilibrium effects are not mixed.


The new ingredient is the contribution of temperature-enhanced scatterings, modeled by the additional term~\ref{eq:delV}
with $c>0$ and $p<0$.
The negative power $p$ captures the fact that scatterings become stronger as the temperature decreases, effectively increasing the thermal mass term for $\phi$ at lower temperatures.
This phenomenological form allows us to explore a wide class of possible microscopic realizations in a model-independent way. \footnote{From a field-theoretic viewpoint, temperature-enhanced scatterings effectively modify the scalar two-point function through the thermal self-energy $\Pi_\phi(T)$. 
When the relevant interaction cross section scales as $\langle\sigma v\rangle\!\propto\!T^{p}$, the loop integral contributing to $\Pi_\phi(T)$ develops an IR enhancement $\Pi_\phi(T)\!\sim\!c\,T^{p}$ for $p<0$. 
This leads to an additional quadratic term $c\,T^{p}\phi^{2}$ in the finite-temperature potential, which parametrizes the dominant low-momentum contribution of such scatterings in a model-independent way.}

At high temperatures, this term is subdominant compared to the usual $a\,T^2\phi^2$ thermal correction (with $a>0$), but as the Universe cools, the term $c\,T^{p}\phi^2$ can become significant, potentially delaying or strengthening the phase transition.

Analytic intuition can be gained by expanding near the origin:
\begin{equation}
V_{\rm eff}(\phi,T)\simeq 
\tfrac12 m_{\rm eff}^2(T)\phi^2 - E T\phi^3 + \tfrac14\lambda\phi^4,
\qquad 
m_{\rm eff}^2(T)=-\mu^2+aT^2+cT^{p}.
\end{equation}
For negative $p$, the $cT^{p}$ term grows relative to $aT^2$ as $T$ decreases, delaying the sign flip of $m_{\rm eff}^2(T)$ and thus increasing supercooling.
This lowers $T_n$, increases the released latent heat $\Delta\rho$, and therefore enhances the strength parameter $\alpha$ while reducing $\beta/H$, in agreement with our numerical results.

\begin{figure}[t]
\centering
\includegraphics[width=0.9\textwidth]{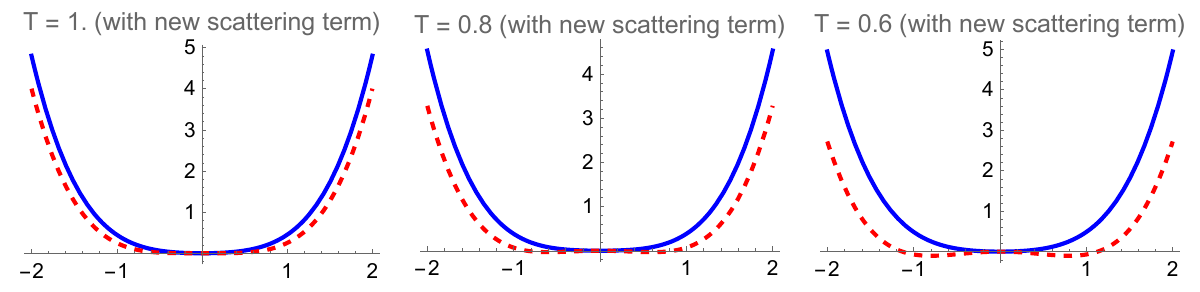}
\caption{
Evolution of the effective potential \(V_{\mathrm{eff}}(\phi,T)\) as the Universe cools, shown at several temperatures \(T\).
Solid lines include the temperature-enhanced scattering term \(c T^p \phi^2\), while dashed lines show the standard potential without this term.
The new term deepens the broken phase minimum or delays the transition, illustrating how temperature-enhanced scatterings can impact the dynamics of the phase transition.
}
\label{fig:VeffPlot}
\end{figure}

Figure~\ref{fig:VeffPlot} illustrates the evolution of the effective potential $V_{\mathrm{eff}}(\phi,T)$ at several representative temperatures.
Solid lines include the temperature-enhanced scattering term, while dashed lines show the standard potential without this new contribution.
As the temperature decreases, the presence of the temperature-enhanced scattering term deepens the broken phase minimum and can modify the critical temperature and nucleation temperature of the phase transition.

In Fig.~\ref{fig:VeffPlot}, the temperatures \(T=1.0, 0.8, 0.6\) are given in dimensionless units scaled by a reference temperature \(T_{\mathrm{ref}}\), which can be interpreted as the characteristic energy scale of the model (for example, the zero-temperature vacuum expectation value of the scalar field or the critical temperature in the absence of the temperature-enhanced term). These values represent a snapshot of the cosmological evolution as the Universe cools down. At the highest temperature \(T=1.0\), the effective potential favors the symmetric phase with the vacuum at \(\phi=0\). As the temperature decreases to \(T=0.8\), the broken minimum at \(\phi \neq 0\) begins to form but may still be metastable. At the lowest temperature shown, \(T=0.6\), the broken phase minimum becomes the global minimum, indicating that the phase transition has completed or is about to complete. The inclusion of the temperature-enhanced scattering term alters the depth and location of these minima, which can significantly affect the critical and nucleation temperatures as well as the strength of the phase transition.
This modification of the effective potential will have a direct impact on the phase transition dynamics and GW signals, as discussed in subsequent sections.

\subsubsection{Explicit thermal-QFT example: one-loop derivation of a $cT^{p}\phi^2$ term}
\label{subsubsec:qft_derivation}

To demonstrate explicitly how temperature-enhanced scatterings can generate an effective quadratic operator of the form $cT^{p}\phi^2$ in the finite-temperature potential, we present a concrete microscopic example and compute the relevant thermal self-energies. The calculation below follows standard finite-temperature field theory (Matsubara) techniques; useful references are \cite{Quiros:1999jp}.

Consider the scalar field \(\phi\) (the order parameter) coupled to a mediator scalar \(S\) and a fermion \(\psi\) in the thermal bath. The interaction Lagrangian is
\begin{equation} \label{eq:UVlag}
\mathcal{L}_{\rm int} \;=\; -\frac{1}{2} m_S^2 S^2 - \mu\, S \,\phi^2 - y_S \, S\,\bar\psi\psi + \mathcal{L}_{\psi,\rm kin},
\end{equation}
where \(\mu\) is a trilinear coupling and \(y_S\) is a Yukawa coupling between the mediator and the bath fermion.\footnote{Other UV completions (for example a direct quartic \(\lambda_{\phi\chi}\phi^2\chi^2\) coupling) lead to related but quantitatively different scalings; the present choice cleanly illustrates the IR-enhancement mechanism discussed in the text.}

The $\phi$ two-point function receives a contribution from the exchange of the mediator \(S\) dressed by fermion loops. In the static, zero-momentum limit the leading contribution to the $\phi$ thermal mass is
\begin{equation} \label{eq:phi_self}
\delta m_\phi^2(T)\;=\;\mu^2\, D_S(p_0=0,\mathbf p\to0;T),
\end{equation}
where \(D_S^{-1}(k_0,\mathbf k;T)=k_0^2+\mathbf k^2+m_S^2+\Pi_S(k_0,\mathbf k;T)\) is the full (thermal) inverse propagator of \(S\) and \(\Pi_S\) is the \(S\)-self-energy arising from the fermion loop. In the local (infrared) approximation one may evaluate \(D_S\) at vanishing external momentum; this yields the effective local operator \(\tfrac12\delta m_\phi^2(T)\phi^2\) in the free-energy.

At one loop the mediator self-energy from the fermion \(\psi\) is (Matsubara sum form)
\begin{equation} \label{eq:PiS_mats}
\Pi_S(0,\mathbf 0;T) \;=\; - y_S^2 \, T \sum_{n\in\mathbb{Z}} \int \!\frac{d^3\mathbf{k}}{(2\pi)^3}\;
\mathrm{Tr}\Big[\,G_\psi(i\omega_n,\mathbf k)\,G_\psi(i\omega_n,\mathbf k)\,\Big],
\end{equation}
where \(G_\psi(i\omega_n,\mathbf k) = {1}/{(i\omega_n \gamma^0 - \boldsymbol{\gamma}\cdot\mathbf k - m_\psi)}\) is the fermion propagator and \(\omega_n=(2n+1)\pi T\) are fermionic Matsubara frequencies. Performing the Dirac trace and the Matsubara sum leads to the well-known result (for a Dirac fermion) in the high-temperature / massless-fermion limit\footnote{We note that more general expressions (including finite fermion mass) can be written using the thermal integral representations; the qualitative behaviour relevant for our argument is captured by the displayed limiting form.}:
\begin{equation} \label{eq:PiS_highT}
\Pi_S(0,\mathbf 0;T) \;\simeq\; \frac{y_S^2}{6}\,T^2 \qquad (m_\psi \ll T).
\end{equation}
This result is textbook material (see e.g. \cite{Quiros:1999jp}) and corresponds to the thermal screening of the scalar mediator due to the bath fermions.

Using \eqref{eq:phi_self} and \eqref{eq:PiS_highT} we obtain
\begin{equation} \label{eq:delta_mphi_full}
\delta m_\phi^2(T) \;=\; \frac{\mu^2}{m_S^2 + \Pi_S(0,\mathbf0;T)}
\;\simeq\; \frac{\mu^2}{m_S^2 + \tfrac{y_S^2}{6}T^2 }.
\end{equation}
Equation \eqref{eq:delta_mphi_full} is the central, explicit result for this minimal model. It is exact at one-loop if we evaluate \(\Pi_S\) in the static limit and then perform the algebraic substitution in the dressed propagator. Eq.~\eqref{eq:delta_mphi_full} demonstrates that the $\phi$ thermal mass is controlled by the ratio \(\mu^2/(m_S^2+\mathcal{O}(y_S^2T^2))\) and hence exhibits distinct asymptotic behaviours:

\begin{itemize}
\item \textbf{Mediator-dominated regime:} If \(m_S^2 \gg y_S^2 T^2\), then \(\delta m_\phi^2(T)\simeq \mu^2/m_S^2\) is approximately temperature independent (corresponding to \(p\simeq 0\) in our phenomenological parametrization).
\item \textbf{Thermally-dressed mediator (IR-enhanced) regime:} If \(y_S^2 T^2 \gg m_S^2\), then
\[
\delta m_\phi^2(T)\simeq \frac{6\mu^2}{y_S^2}\,T^{-2},
\]
i.e. \(\delta m_\phi^2(T)\propto T^{-2}\) (so \(p=-2\) in the parametrization \(cT^{p}\)). This negative power arises because the fermion-loop dressing of the mediator produces a thermal screening mass \(\Pi_S\propto T^2\) in the denominator of the dressed propagator, causing the overall $\phi$ mass correction to scale as the inverse of \(T^2\).
\end{itemize}

Thus, in this explicit model one obtains a concrete realization of \(cT^{p}\phi^2\) with \(p\le 0\) depending on the parametric hierarchy between \(m_S\) and \(y_S T\). Importantly, the negative power \(p=-2\) occurs naturally in the regime where the mediator is light compared to the thermal scale but acquires a thermal self-energy that grows like \(T^2\).

Although the derivation above used the dressed-mediator picture, the same physical origin can be viewed from the perspective of 2\(\to\)2 scatterings in the thermal bath: \(\psi\phi\to\psi\phi\) proceeds dominantly via \(t\)-channel \(S\)-exchange with amplitude \(\mathcal{M}\sim \mu y_S/(t-m_S^2)\). In the thermal average the typical momentum transfer satisfies \(|t|\sim T^2\), so the squared amplitude scales as \(|\mathcal{M}|^2\propto \mu^2 y_S^2/(T^2+m_S^2)^2\). Thermally averaging and integrating over phase space yields an effective rate with inverse powers of \(T\) in regimes where \(m_S\) is small; the dressed-mediator self-energy computation above captures this effect in equilibrium as a modification of the static propagator and therefore as a \(\phi^2\)-type contribution to the effective potential. Both viewpoints are consistent: the equilibrium self-energy is the field-theoretic image of forward (coherent) scatterings with the thermal bath.

The example shows that \(p\) is model-dependent and can take values \(p\le 0\) depending on the microscopic spectrum and couplings. For the toy model \eqref{eq:UVlag} we found \(p=0\) (heavy mediator) or \(p=-2\) (light mediator with thermal dressing). Other microscopic scenarios with more infrared-divergent kernels (for example, additional long-range interactions or multiple-step dressing) can produce stronger inverse-temperature scalings in restricted parameter regions; however the dressed-propagator argument above shows that the appearance of a quadratic-in-\(\phi\) thermal operator is generic, while the precise exponent \(p\) is fixed by the dominant thermal dressing mechanism.

\vspace{6pt}
\noindent\textbf{Connection with the effective potential:}
Working in the Matsubara formalism, the static (zero-frequency, zero-momentum) limit of the one-particle-irreducible two-point function is precisely the second derivative of the free-energy functional with respect to the field and thus corresponds to a quadratic term in \(V_{\rm eff}\):
\[
\Pi_\phi(0,0;T) \;=\; \left.\frac{\partial^2 V_{\rm eff}(\phi,T)}{\partial\phi^2}\right|_{\phi=0}.
\]
The replacement \(\tfrac12\Pi_\phi(T)\phi^2\) in \(V_{\rm eff}\) is therefore exact at the level of the local, infrared-dominated contribution. This is why the equilibrium self-energy calculation is the appropriate framework to incorporate temperature-dependent scattering effects into the static effective potential used to compute nucleation and tunneling. Non-equilibrium collision terms (Boltzmann integrals) govern transport and friction; they affect wall dynamics but do not replace the equilibrium thermal mass computation above.

\vspace{6pt}
\noindent\textbf{Caveats and practical implementation:}
\begin{itemize}
\item The relation \eqref{eq:delta_mphi_full} was derived in the static limit and at one-loop order. It captures the leading equilibrium contribution relevant for the effective potential and nucleation computation. Higher-loop corrections and non-local momentum-dependent pieces can modify the numerical coefficients and produce subleading corrections to the shape of the potential; these effects can be incorporated systematically if required.
\item For numerical work we insert the parametrized form \(\delta m_\phi^2(T)=cT^{p}\) into \(V_{\rm eff}\). The constant \(c\) and exponent \(p\) should be chosen (or matched) according to the microscopic relations such as \eqref{eq:delta_mphi_full}; for the toy model the mapping is \(c\simeq 6\mu^2/y_S^2\) and \(p=-2\) in the thermally-dressed regime.
\item The derivation above demonstrates that the extra term is an equilibrium self-energy effect and therefore is not double-counted if the mediator and its thermal dressing are not already included in the baseline thermal free-energy \(V_T\) used in the numerics.
\end{itemize}

\subsection{Phase transition parameters}
\label{subsec:parameters}

The dynamics of a cosmological first-order phase transition are typically characterized by several key parameters~\cite{Caprini:2015zlo, Caprini:2019egz, Mazumdar:2018dfl, Hindmarsh:2020hop, Athron:2023mer}:
\begin{itemize}
\item \textbf{Critical temperature} $T_c$: the temperature at which the symmetric and broken minima of the thermal effective potential $V_{\mathrm{eff}}(\phi,T)$ become degenerate~\cite{Quiros:1999jp, Dine:1992wr, Cline:2006ts}.
\item \textbf{Nucleation temperature} $T_n$: the temperature at which bubbles of the broken phase nucleate efficiently, defined by the condition $S_3(T_n)/T_n \approx 140$, where $S_3(T)$ is the three-dimensional Euclidean bounce action~\cite{Linde:1980tt, Andersen:2004fp, Quiros:1999jp}. This ensures at least one bubble per Hubble volume is nucleated.
\item \textbf{Strength parameter} $\alpha$: the ratio of the latent heat released during the phase transition to the radiation energy density at $T_n$, which controls the fraction of vacuum energy converted into bulk motion and GW~\cite{Caprini:2015zlo, Grojean:2006bp, Espinosa:2010hh}.
\item \textbf{Inverse duration parameter} $\beta$: defined as $\beta/H  \big|_{T_n}$, where $H$ is the Hubble parameter at $T_n$~\cite{Caprini:2015zlo, Caprini:2019egz}. It measures how rapidly the transition completes; smaller $\beta/H$ corresponds to a slower, longer transition producing stronger GW signals.
\end{itemize}

In our framework, temperature-enhanced scatterings introduce an additional term of the form $c\, T^{p} \phi^2$ in the effective potential, with $p<0$~\cite{Baldes:2018nel, Bian:2019bsn, Athron:2023mer}. 
As the Universe cools, this term becomes increasingly significant, modifying the thermal evolution of $V_{\mathrm{eff}}(\phi,T)$ and directly affecting the bounce action $S_3(T)/T$, where $S_3(T)$ is the three dimensional Euclidean action. Details about the bounce action is given in Appendix \ref{app:bounce}.
As illustrated in Fig.~\ref{fig:BounceAction}, different values of $p$ change the temperature dependence of $S_3(T)/T$, shifting the nucleation temperature $T_n$, altering the duration of the phase transition (through $\beta/H$), and affecting its strength (through $\alpha$).

\begin{figure}[t]
\centering
\includegraphics[width=0.75\textwidth]{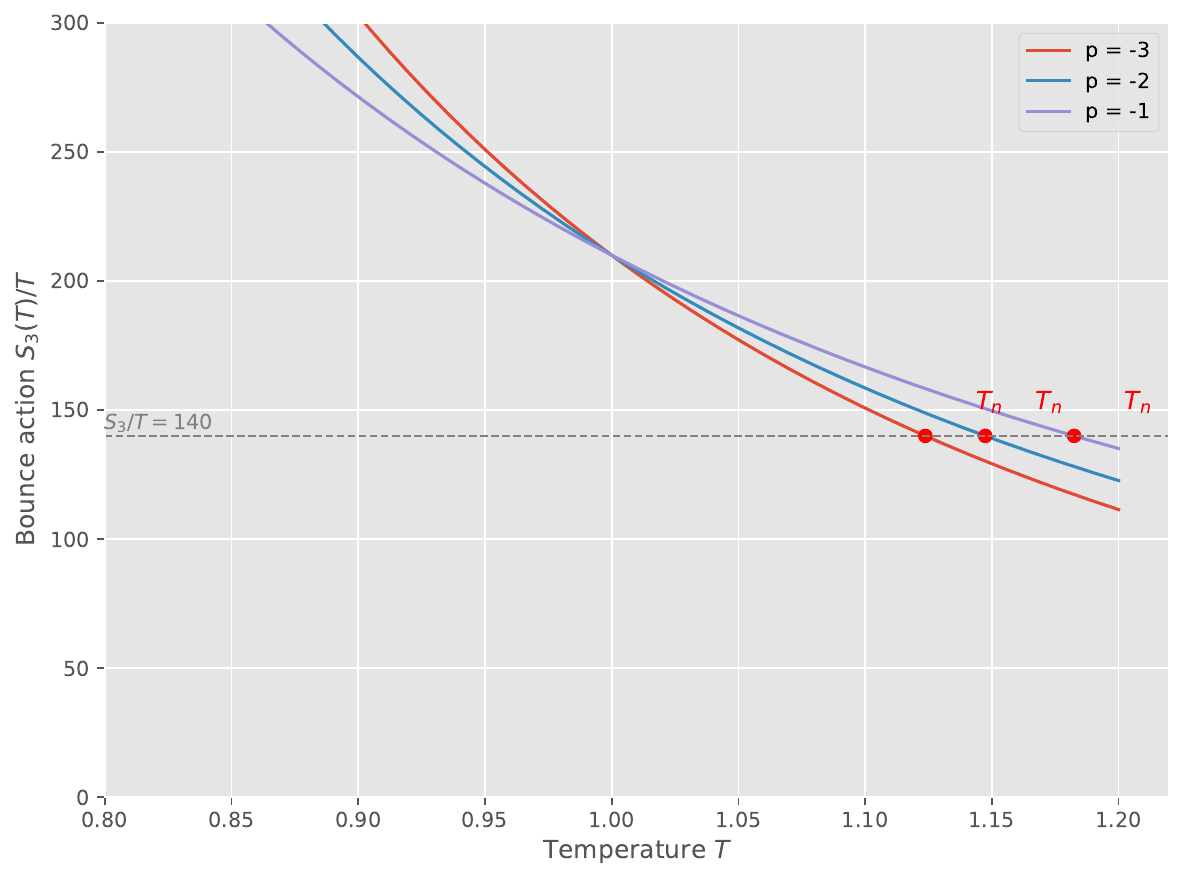}
\caption{
Temperature dependence of the bounce action \(S_3(T)/T\) for different exponents \(p\) in the temperature-enhanced scattering term. 
The horizontal dashed line at \(S_3/T=140\) indicates the nucleation criterion.
Red points mark the nucleation temperature \(T_n\) where the bounce action drops below this threshold.
The temperature on the x-axis is shown in units of the critical temperature, \(T/T_c\), making the plot dimensionless.
As \(p\) varies, the dynamics of the phase transition change, shifting the nucleation temperature and affecting the strength and duration of the transition.
}
\label{fig:BounceAction}
\end{figure}

For analytic intuition, one can approximate the effective potential near the origin as $V_{\mathrm{eff}}\!\approx\!\tfrac12 m_{\mathrm{eff}}^{2}(T)\phi^{2}-E T\phi^{3}+\tfrac14\lambda\phi^{4}$, where $E$ collects cubic thermal contributions.  A negative $p$ increases the magnitude of $m_{\mathrm{eff}}^{2}(T)$ at low $T$. Thereby deepening the broken minimum and lowering $T_c$ and $T_n$.  This qualitative behavior explains the trends seen in Fig.~\ref{fig:BounceAction}.

These phase transition parameters directly determine the amplitude, peak frequency, and shape of the stochastic GW background produced in the early Universe~\cite{Caprini:2015zlo, Caprini:2019egz, Athron:2023mer}.
In the following sections, we numerically construct $V_{\mathrm{eff}}(\phi,T)$ including temperature-enhanced scatterings, extract these parameters, and explore their phenomenological consequences for GW signals.

\subsection{Real-Singlet Extension as a Concrete UV Realization}
\label{sec:UVexampledetail}

To illustrate how a temperature-dependent quadratic correction of the form
$c\,T^{p}\phi^{2}$ naturally emerges in a realistic particle-physics model, we
briefly summarize the relevant features of the real-singlet extension of the
Standard Model (xSM).  This minimal scalar extension is well known to support a
strong first-order electroweak phase transition through a tree-level barrier,
while simultaneously generating thermal self-energy corrections to the Higgs
field that match the structure of Eq.~\eqref{eq:delV}.

\subsubsection*{Lagrangian and mixing structure}
The scalar potential of the $Z_{2}$-symmetric xSM is
\begin{equation}
V(H,S)
= -\mu_{H}^{2} |H|^{2} + \lambda_{H} |H|^{4}
  + \frac{1}{2}\mu_{S}^{2} S^{2} + \frac{1}{4}\lambda_{S} S^{4}
  + \frac{1}{2}\lambda_{HS} |H|^{2} S^{2},
\label{eq:xSMpot}
\end{equation}
where $H$ is the Higgs doublet and $S$ is a real gauge singlet.
After electroweak symmetry breaking, the fields mix due to the portal
interaction $\lambda_{HS} |H|^{2}S^{2}$, producing the mass eigenstates
$h_{1}$ and $h_{2}$ with mixing angle $\theta$.

\subsubsection*{Thermal self-energy and mapping to \texorpdfstring{$c\,T^{p}, \phi^2$}{cT^p\phi^2}}
The one-loop finite-temperature Higgs self-energy receives a contribution from
singlet loops,
\begin{equation}
\Pi_{h}(T) \;\supset\;
\frac{\lambda_{HS}}{12}\,T^{2}\; F(m_{S}/T),
\label{eq:xSMselfenergy}
\end{equation}
where $F(x)$ captures the deviation from the pure high-temperature limit and
interpolates between $F(x)\!\to\!1$ for $x\ll 1$ and
$F(x)\!\propto\! x e^{-x}$ for $x\gg 1$.  In the regime where the singlet is
non-relativistic during the phase transition, $m_{S}\gtrsim T$, the Boltzmann
suppression produces a temperature dependence that is well approximated by
a power law,
\begin{equation}
F(m_{S}/T) \;\simeq\; A\,T^{\,p},
\qquad p<0,
\end{equation}
with the exponent determined by the dominant microscopic scattering channels.
Inserting this into Eq.~\eqref{eq:xSMselfenergy} and using the local
approximation for the one-particle-irreducible effective potential yields
\begin{equation}
\Delta V_{\text{xSM}}(\phi,T)
= \Pi_{h}(T)\,\frac{\phi^{2}}{2}
\simeq
c\,T^{p}\phi^{2},
\label{eq:xSMmapping}
\end{equation}
where the matching coefficient is
\begin{equation}
c \;=\; \frac{\lambda_{HS}}{24}\,A .
\end{equation}
Thus the xSM provides a concrete UV origin for both the sign and the magnitude
of the $(c,p)$ parameters used in our phenomenological analysis.

\subsubsection*{Phase-transition relevance}
The portal interaction $\lambda_{HS}$ also generates a tree-level barrier in
the scalar potential when the singlet acquires a thermal expectation value or
develops temperature-dependent mass corrections.  Numerous studies have shown
that for natural parameter choices, the xSM supports a strongly first-order
electroweak phase transition consistent with Higgs-coupling constraints.  In
such regions of parameter space, the thermal self-energy term in
Eq.~\eqref{eq:xSMmapping} constitutes an additional, dynamically evolving
contribution to the curvature of the effective potential, and therefore fits
squarely into the general framework developed in this paper.

This example demonstrates explicitly that the parametrization in
Eq.~\eqref{eq:delV} is not merely phenomenological, but directly realizable in
well-motivated extensions of the Standard Model.  In the remainder of this
work, we focus on the $(c,p)$ description, which efficiently captures the
leading impact of such thermal dressing effects on phase-transition dynamics
and the resulting gravitational-wave signal.

We emphasize that the assumption $m_S \gtrsim T$ does not conflict with the presence of a
tree-level barrier or the realization of a first-order phase transition.
In singlet-extended scalar theories, the barrier responsible for the phase transition
originates from the zero-temperature structure of the scalar potential and, where relevant,
from light degrees of freedom whose thermal functions admit a high-temperature expansion.
In contrast, the singlet mediator $S$ responsible for the temperature-dependent mass correction
$\delta m_\phi^2(T)$ need not be light and is treated separately through its static thermal
self-energy.
The heavy-mediator regime $m_S \gtrsim T$ therefore applies only to the sector generating the
effective $c\,T^{p}\phi^2$ contribution and does not affect the validity of the tree-level
barrier or the phase transition dynamics.
Accordingly, no assumption of a global high-temperature expansion is made, and different
thermal regimes are consistently applied to different sectors of the theory.

\section{Gravitational wave prediction}
\label{sec:gw}

\subsection{GW sources from FOPTs}
\label{subsec:sources}

A FOPT in the early Universe generates a stochastic background of GWs through several distinct physical processes. The primary sources include bubble collisions, sound waves, and magnetohydrodynamic (MHD) turbulence, each contributing uniquely to the overall GW spectrum~\cite{Caprini:2015zlo, Caprini:2019egz, Hindmarsh:2020hop}.

Bubble collisions arise from the dynamics of expanding true vacuum bubbles that nucleate and collide during the phase transition. These collisions produce GWs primarily through scalar field gradients localized at the bubble walls, as initially studied in Refs.~\cite{Kosowsky:1992rz, Kosowsky:1992vn}. Although this mechanism can yield a distinct spectral shape, its contribution is often subdominant compared to fluid effects in many realistic scenarios.

The dominant source in most cosmological first-order phase transitions is the sound waves generated by the coherent bulk motion of the plasma. As expanding bubbles push the surrounding plasma, sound waves propagate and persist over timescales comparable to the Hubble time, efficiently sourcing gravitational radiation~\cite{Hindmarsh:2013xza, Hindmarsh:2015qta}. The resulting GW signal from sound waves typically has a characteristic spectral peak and amplitude that depend sensitively on the strength and duration of the phase transition.

In addition, turbulent motion of the plasma and the associated magnetic fields induced after bubble collisions produce GWs through magnetohydrodynamic (MHD) turbulence~\cite{Caprini:2006jb, Caprini:2009yp}. Although turbulence is expected to be a subdominant source compared to sound waves, it can contribute non-negligibly to the GW spectrum at high frequencies and alter the spectral shape.

For clarity, the total GW spectrum is expressed as a sum of three components,
\begin{align}
\Omega_{\mathrm{GW}}(f) h^2 &= \Omega_{\mathrm{col}}(f) h^2 + \Omega_{\mathrm{sw}}(f) h^2 + \Omega_{\mathrm{turb}}(f) h^2, \label{eq:omega_total}
\end{align}
with explicit forms given below~\cite{Caprini:2019egz, Huber:2008hg, Hindmarsh:2017gnf}:
\begin{align}
\Omega_{\mathrm{col}}(f) h^2 &= 1.67\times10^{-5}
\left( \frac{H}{\beta} \right)^2
\left( \frac{\kappa_{\phi}\,\alpha}{1+\alpha} \right)^2
\left( \frac{100}{g_*} \right)^{1/3}
\frac{0.11\,v_w^3}{0.42+v_w^2}
S_{\mathrm{env}}\!\left(\frac{f}{f_{\mathrm{col}}}\right), \\[4pt]
\Omega_{\mathrm{sw}}(f) h^2 &= \Upsilon\cdot 2.65\times10^{-6}
\left( \frac{H}{\beta} \right)
\left( \frac{\kappa_{\mathrm{sw}}\,\alpha}{1+\alpha} \right)^2
\left( \frac{100}{g_*} \right)^{1/3}
v_w\,S_{\mathrm{sw}}\!\left(\frac{f}{f_{\mathrm{sw}}}\right), \\[4pt]
\Omega_{\mathrm{turb}}(f) h^2 &= {\cal O}(1)\times 3.35\times10^{-4}
\left( \frac{H}{\beta} \right)
\left( \frac{\kappa_{\mathrm{turb}}\,\alpha}{1+\alpha} \right)^{3/2}
\left( \frac{100}{g_*} \right)^{1/3}
v_w\,S_{\mathrm{turb}}\!\left(\frac{f}{f_{\mathrm{turb}}}\right),
\end{align}
where $S_i(f/f_i)$ denote normalized spectral shapes for each source, and the efficiency factors $\kappa_i(\alpha,v_w)$ represent the fraction of released vacuum energy converted into the corresponding channel. Among these, the sound-wave contribution typically dominates the integrated signal for subluminal wall velocities, while turbulence adds a high-frequency tail.
We use the semi-analytic fits summarized above following the review~\cite{Caprini:2019egz} and\cite{Huber:2008hg} (bubble collisions), and the sound-shell model results of \cite{Hindmarsh:2017gnf}.  Recent numerical simulations indicate that the acoustic source can decay on timescales shorter than a Hubble time for sufficiently strong transitions; to account for this effect we multiply \(\Omega_{\rm sw}\) by a suppression factor \(\Upsilon\) (typically \(\Upsilon=\min(1,H\tau_{\rm sw})\) or a fit derived from simulations).  The turbulence contribution remains subject to larger modeling uncertainties and should be regarded as an order-of-magnitude estimate.

The overall shape, peak frequency, and amplitude of the stochastic GW background generated by a first-order phase transition are controlled by the key parameters characterizing the transition: the strength parameter \(\alpha\), which measures the ratio of vacuum energy released to the radiation energy density; the inverse duration parameter \(\beta/H\), which quantifies the rapidity of the transition; the bubble wall velocity \(v_w\); and the nucleation temperature \(T_n\). In our framework, temperature-enhanced scatterings modify the thermal effective potential and thus affect the evolution of the bounce action. This leads to changes in these phase transition parameters, ultimately influencing the predicted GW signal.

The dominant contribution in our analysis comes from sound waves, for which the peak frequency in Eq.~(8) corresponds to $f_{\mathrm{sw}}$. This is justified because the plasma motion persists for roughly a Hubble time after the transition, making the sound-wave source more efficient than the instantaneous envelope contribution from bubble collisions.

In the following subsection, we detail the computational approach employed to quantitatively evaluate the contributions from these sources and to predict the resulting GW spectra.

\subsection{Computation setup}
\label{subsec:setup}

To compute the stochastic GW spectrum generated by the first-order phase transition, we adopt the standard semi-analytic formalism developed in Refs.~\cite{Caprini:2015zlo, Caprini:2019egz, Huber:2008hg, Hindmarsh:2015qta}. In this framework, the total GW energy density spectrum today can be expressed as the sum of three main contributions, as shown in Eq.~\ref{eq:omega_total}.

The shape and peak frequency of each component depend on the thermal parameters characterizing the phase transition. The peak frequency today can be approximately written as
\begin{equation}
f_{\mathrm{peak}} \simeq 1.65 \times 10^{-5}~\mathrm{Hz} \left( \frac{f_*}{\beta} \right) \left( \frac{\beta}{H} \right) \left( \frac{T_n}{100\,\mathrm{GeV}} \right) \left( \frac{g_*}{100} \right)^{1/6}.
\end{equation}
This expression corresponds to the sound-wave source, with $f_* / \beta \simeq 0.9 /(1.8-0.1v_w+v_w^2)$ as given in~\cite{Caprini:2019egz}. For bubble collisions and turbulence, analogous relations hold with slightly shifted prefactors ($f_* / \beta \simeq 0.62/(1.8-0.1v_w+v_w^2)$ for the envelope approximation and $f_* / \beta \simeq 1.4/(1+v_w)$ for turbulence).

The amplitude of each contribution is controlled by the strength parameter \(\alpha\), the bubble wall velocity \(v_w\), and the efficiency factors \(\kappa_i(\alpha, v_w)\) which quantify the fraction of the latent heat converted into bulk motion, sound waves, or turbulence. The bubble wall velocity \(v_w\) is treated as a phenomenological parameter, typically varied within \(v_w \in [0.6, 1.0]\). The values of \(\alpha\) and \(\beta/H\) are extracted numerically from the temperature-dependent effective potential \(V_{\mathrm{eff}}(\phi,T)\) described in earlier sections, which includes the temperature-enhanced scattering term \(c\, T^{p} \phi^2\). 

Our computation proceeds by first scanning over the parameters \(p\) and \(c\) to generate a set of effective potentials for different choices of the temperature-enhanced scattering term. For each parameter point, we compute the bounce action \(S_3(T)/T\) as a function of temperature to extract the nucleation temperature \(T_n\). Using \(T_n\), we then determine the strength parameter \(\alpha\) by calculating the released latent heat, and evaluate \(\beta/H\) from the temperature derivative of the bounce action at nucleation:
\begin{equation} 
\beta/H = T_n \left. \frac{d(S_3/T)}{dT} \right|_{T=T_n}.
\end{equation}

All bounce actions and tunneling profiles were computed with the public package \texttt{CosmoTransitions}~\cite{Wainwright:2011kj}, into which we incorporated the $cT^{p}\phi^{2}$ term. \footnote{The implementation modifies only the scalar mass matrix passed
to \texttt{CosmoTransitions}; no additional thermal prefactors or mass resummations
are introduced.}
  For validation, we cross-checked representative points using an independent overshoot/undershoot routine described in Appendix~\ref{app:bounce}, finding agreement at the few-percent level.

The scattering-induced contribution $cT^{p}\phi^{2}$ is incorporated only through the
field-dependent mass in the \emph{tree-level} curvature entering $V_{\rm eff}(\phi,T)$.
The remaining thermal free-energy contributions are computed using the exact
bosonic and fermionic integrals $J_{B,F}(m^{2}/T^{2})$, without employing a
high-temperature expansion. This avoids double counting: the same mediator loop
responsible for $cT^{p}\phi^{2}$ is \textit{not} reinserted into the thermal functions $J_{B,F}$,
where the propagators are already thermally dressed. Consequently, the nucleation
action and thermodynamic parameters $\{\alpha,\,\beta/H,\,T_n\}$ are evaluated from a
potential in which the scattering correction appears \emph{once and only once}, in
agreement with the static-limit derivation of Appendix~\ref{app:matsubara}.

Once these thermal parameters are known, individual contributions \(\Omega_{\mathrm{col}}\), \(\Omega_{\mathrm{sw}}\), and \(\Omega_{\mathrm{turb}}\) to the GW spectrum are evaluated using the semi-analytic expressions above. This approach captures the essential physics of GW generation from expanding and colliding bubbles, as well as the subsequent sound waves and turbulence in the plasma.

The explicit spectral shapes and efficiency factors used in our computation are as follows:
\begin{align}
S_{\mathrm{env}}(x) &= \frac{3.8\,x^{2.8}}{1+2.8\,x^{3.8}}, \quad \text{(envelope approximation)~\cite{Huber:2008hg}}, \\
S_{\mathrm{sw}}(x) &= \frac{(7x)^{6.8}}{6.8x^{1.6}+1.36x^{3.8}+3.8x^{7.0}}, \quad \text{(sound-wave fit)~\cite{Hindmarsh:2017gnf, Caprini:2019egz}}, \\
S_{\mathrm{turb}}(x) &= \frac{x^{3}}{(1+x)^{11/3}\left(1+8\pi f/h_*\right)}, \quad \text{(MHD turbulence fit)~\cite{Caprini:2009yp}}.
\end{align}

The efficiency factors converting released vacuum energy into fluid motion are taken from~\cite{Espinosa:2010hh}:
\begin{align}
\kappa_{\mathrm{sw}}(\alpha,v_w) &\simeq \frac{\alpha}{0.73+0.083\sqrt{\alpha}+\alpha}, \\
\kappa_{\phi}(\alpha,v_w) &\simeq \frac{0.715\,\alpha+4\sqrt{3\alpha/2}}{1+0.715\,\alpha}, \\
\kappa_{\mathrm{turb}} &= 0.1\,\kappa_{\mathrm{sw}}.
\end{align}
We include the acoustic suppression factor $\Upsilon$~\cite{Hindmarsh:2020hop},
\begin{equation}
\Upsilon = 1 - \frac{1}{\sqrt{1 + 2\,H\,\tau_{\mathrm{sw}}}}, \quad
\tau_{\mathrm{sw}} \simeq \frac{R_*}{U_f},
\end{equation}
where $R_*$ is the mean bubble separation and $U_f$ is the root-mean-square fluid velocity.
We apply $\Upsilon$ as a multiplicative damping of $\Omega_{\mathrm{sw}}$ to account for the finite lifetime of the acoustic source.

In all numerical plots, the total spectrum is the sum of the three components,
$\Omega_{\mathrm{GW}} = \Omega_{\mathrm{col}} + \Omega_{\mathrm{sw}} + \Omega_{\mathrm{turb}}$,
with sound waves dominating ($\gtrsim90\%$ of the total signal), followed by bubble collisions ($\lesssim10\%$) and turbulence ($\lesssim5\%$).
We take the turbulence prefactor $3.35\times10^{-4}$ as calibrated in~\cite{Caprini:2019egz} and assume $v_w=0.9$ for the benchmark spectra unless stated otherwise.
These prescriptions reproduce the spectral shapes observed in numerical simulations and ensure direct comparability with existing literature.

Throughout this analysis, we systematically vary the exponent \(p\) and the coefficient \(c\) to explore how temperature-enhanced scatterings affect the key phase transition parameters and, consequently, the predicted GW spectra. The goal is to identify regions of parameter space where the resulting GW signals fall within the projected sensitivity ranges of future experiments such as LISA, DECIGO, and BBO.

\subsection{Bubble wall friction estimate}

The motion of expanding bubble walls during a first-order phase transition is subject to friction due to interactions with particles in the thermal plasma~\cite{Espinosa:2010hh}. In our framework, the temperature-enhanced scatterings can modify this friction by increasing the effective interaction rate between the scalar field $\varphi$ and thermal bath particles.

At leading order, the friction coefficient $\eta$ is proportional to the thermal interaction rate,
\begin{equation}
    \eta \sim \sum_{i} \Delta m_i^2\, n_i(T) / \Gamma_i,
\end{equation}
where $\Delta m_i^2$ is the change in the particle’s mass squared across the bubble wall, $n_i(T)$ is the thermal number density, and $\Gamma_i$ is the interaction rate. In the presence of enhanced scatterings, the rate $\Gamma_i$ increases due to the $T$-dependent cross section $\sigma(T) \propto T^{-n}$.

Consequently, the friction grows as the Universe cools. However, for the range of parameters we consider ($n \sim 1{-}4$), the friction enhancement remains moderate and does not prevent bubble expansion, provided the latent heat is sufficient to accelerate the wall. 

To ensure that our semi-analytic treatment remains consistent, we have verified that the bubble wall velocity $v_w$ remains in the typical range $0.6 \lesssim v_w \lesssim 1$, where sound wave production dominates the GW signal. A more detailed calculation of friction and wall dynamics would require a coupled treatment of Boltzmann transport equations and is beyond the scope of this study.

In the following section, we present the detailed numerical results, illustrating the dependence of the GW peak amplitude and frequency on the model parameters and discussing the phenomenological implications for experimental detection.

\section{Parameter space analysis and phenomenology}
\label{sec:parameter_scan}

In this section, we systematically explore the parameter space associated with the temperature-enhanced scattering term in the thermal effective potential,
\begin{equation}
    V_{\mathrm{eff}}(\phi, T) = V_0(\phi) + \Delta V_{\mathrm{thermal}}(\phi, T) + c\, T^p \phi^2,
\end{equation}
where \(c\) is a dimensionless coefficient controlling the strength of the scattering effect, and \(p<0\) determines how this contribution scales with temperature.
The new term captures scatterings whose cross sections increase as the Universe cools, enhancing the phase transition at late times.
By scanning over the ranges
\begin{equation}
    p \in [-1.5, 0], \quad c \in [0, c_{\max}],
\end{equation}
where \(c_{\max}\) is chosen to preserve perturbativity and vacuum stability, we quantify how these parameters affect the dynamics of the electroweak-like phase transition and the resulting GW signatures. 
The upper limit \(c_{\max}\) is fixed by requiring perturbativity and vacuum stability:
(i) $|\Pi_\phi(T)|\lesssim 4\pi v^2$ for $T$ near the transition so that loop corrections remain perturbative, and
(ii) the zero-temperature vacuum stays the global minimum.
For the electroweak-like benchmarks considered, these conditions correspond to $c_{\max}\!\sim\!\mathcal{O}(0.1\!-\!1)$.

The effective scalar mass term in this setup is
\(m_{\mathrm{eff}}^2(T) = -\mu^2 + a\,T^2 + c\,T^{p}\),
with \(a>0\) the usual finite-temperature coefficient from one-loop self-energy diagrams and \(cT^{p}\) the additional correction from temperature-enhanced scatterings.
For \(p<0\), the last term grows in relative importance as the Universe cools, steepening the curvature of the potential near the origin and thereby delaying the symmetry-breaking transition.
Once the barrier separating the symmetric and broken minima disappears, the larger latent heat amplifies the strength parameter \(\alpha\) and prolongs the duration of the transition, leading to smaller \(\beta/H\).

Our analysis proceeds in three stages:
\begin{itemize}
\item compute the temperature-dependent effective potential \(V_{\mathrm{eff}}(\phi,T)\) across the parameter space,
\item extract the key phase transition parameters---critical temperature \(T_c\), nucleation temperature \(T_n\), strength parameter \(\alpha\), and inverse duration parameter \(\beta/H\),
\item predict the stochastic GW signal and assess its observability at upcoming experiments such as LISA, DECIGO, and BBO.
\end{itemize}

\vspace{8pt}
\noindent
The results are summarized below.

\subsection{Scanning the temperature-enhanced scattering parameters}
\label{subsec:scan_scattering}

For each pair \((p,c)\), we numerically compute \(V_{\mathrm{eff}}(\phi,T)\) and track the evolution of its minima with temperature.
The critical temperature \(T_c\) is defined by the condition
\begin{equation}
    V_{\mathrm{eff}}(0,T_c) = V_{\mathrm{eff}}(\phi_{\mathrm{min}}(T_c), T_c),
\end{equation}
where \(\phi_{\mathrm{min}}(T)\) is the nontrivial minimum.
The nucleation temperature \(T_n\) is determined by solving
\begin{equation} \label{eq:Tnuc}
    \frac{S_3(T_n)}{T_n} \approx 140,
\end{equation}
with \(S_3(T)\) the three-dimensional Euclidean bounce action.

We also compute:
\begin{align}
    \alpha &= \frac{\Delta \rho}{\rho_{\mathrm{rad}}(T_n)}, \label{eq:alpha} 
    \\
    \frac{\beta}{H} &= T_n \left. \frac{d}{dT} \left( \frac{S_3(T)}{T} \right) \right|_{T_n}, \label{eq:beta}
\end{align}
where \(\Delta \rho\) is the released latent heat and \(\rho_{\mathrm{rad}}(T_n)\) is the radiation energy density at \(T_n\).

For numerical stability, we impose that the zero-temperature vacuum remains the global minimum and that \(c\,T^{p}\) never drives \(m_{\mathrm{eff}}^2(T)\) negative at high temperature. These criteria constrain \(c_{\max}\) to the range \(0.1 \lesssim c_{\max} \lesssim 1\) for typical benchmark choices, ensuring the potential remains perturbative and stable.

\begin{figure}[t]
    \centering
    \includegraphics[width=0.49\textwidth]{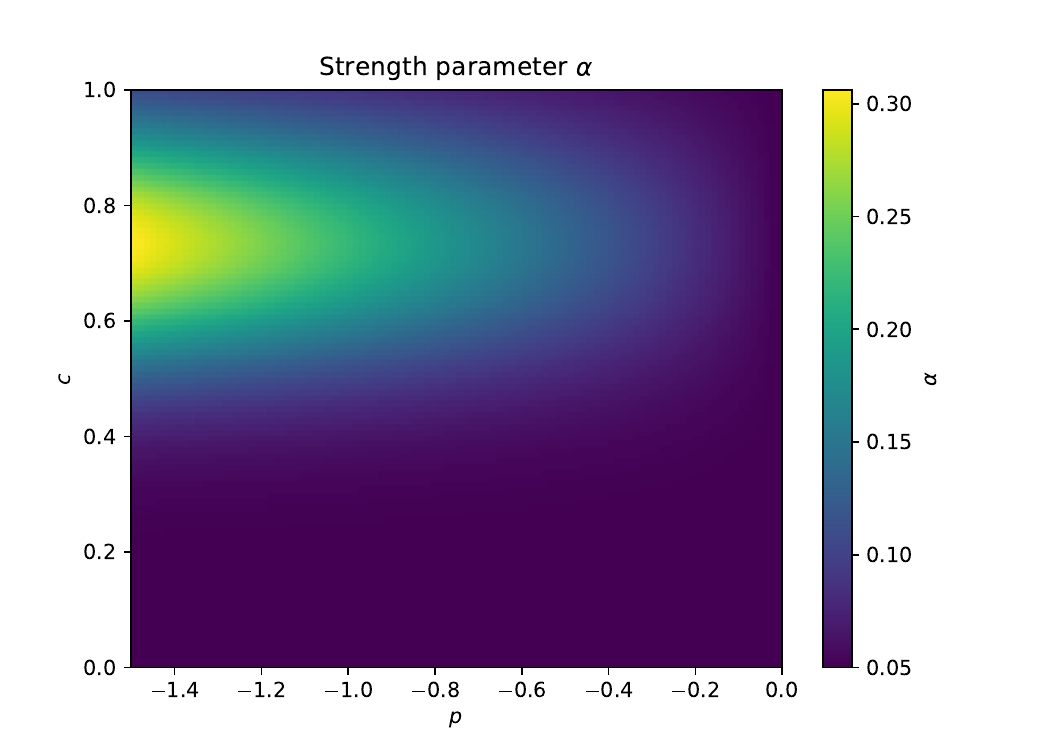}
    \includegraphics[width=0.49\textwidth]{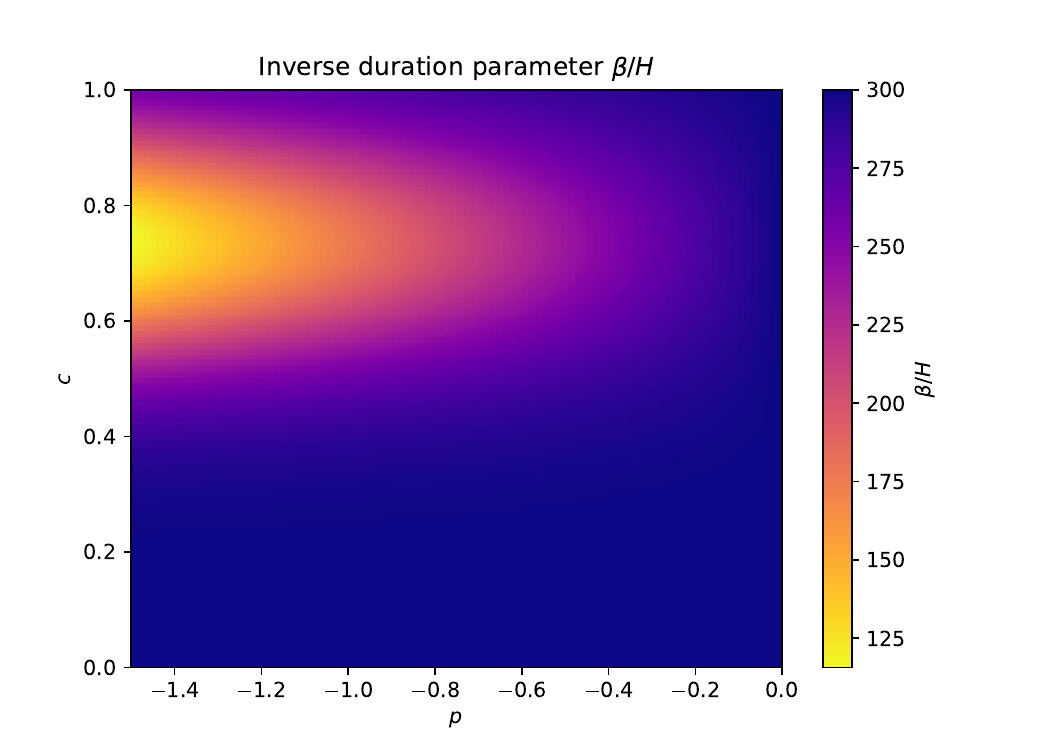}
    \caption{
    Scan over the temperature-enhanced scattering parameters \(p\) and \(c\). \textit{Left:} Strength parameter \(\alpha\). \textit{Right:} Inverse duration parameter \(\beta/H\). 
    Regions with larger \(\alpha\) and smaller \(\beta/H\) correspond to stronger and longer-lasting phase transitions, most pronounced for large negative \(p\) and moderate \(c\).
    }
    \label{fig:phaseparams_scan}
\end{figure}

\vspace{6pt}
\noindent
\textbf{Figure~\ref{fig:phaseparams_scan}} presents the results:
\begin{itemize}
\item \textit{Left:} The strength parameter \(\alpha\) increases significantly as \(p\) becomes more negative and \(c\) reaches moderate values, reflecting enhanced symmetry breaking at lower temperatures.
\item \textit{Right:} The inverse duration parameter \(\beta/H\) correspondingly decreases, indicating slower transitions that last longer, which is favorable for GW production.
\end{itemize}

These behaviors follow directly from the modified thermal mass: a larger negative \(p\) deepens the potential well more sharply near the transition, yielding stronger supercooling and a higher ratio of latent heat to background radiation. The smaller temperature derivative of the bounce action near \(T_n\) reduces \(\beta/H\), lengthening the GW-generating stage.

\subsection{Impact on phase transition parameters}
\label{subsec:scan_phase_transition}

\begin{figure}[h]
    \centering
    \includegraphics[width=0.6\textwidth]{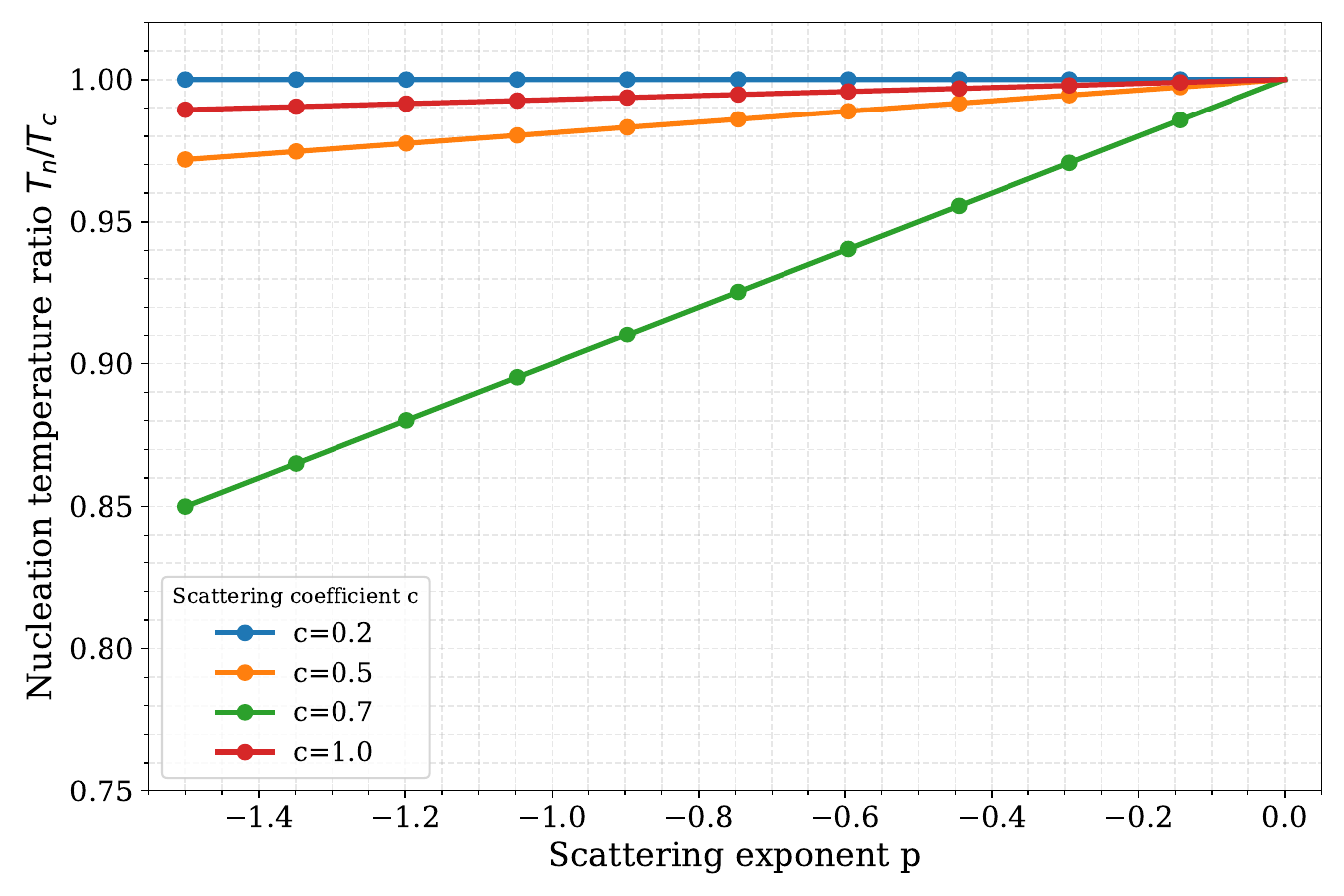}
    \caption{
    Nucleation temperature \(T_n\) normalized to the critical temperature \(T_c\) versus exponent \(p\) for different values of \(c\).
    Larger negative \(p\) leads to later nucleation (lower \(T_n/T_c\)), as the scattering term increasingly favors the broken phase at lower temperatures.
    }
    \label{fig:Tn_scan}
\end{figure}

Beyond \(\alpha\) and \(\beta/H\), the temperature-enhanced scatterings also modify \(T_c\) and \(T_n\). 
Figure~\ref{fig:Tn_scan} shows how \(T_n/T_c\) varies with \(p\) for several benchmark values of \(c\).
As \(p\) becomes more negative, \(T_n\) shifts to lower values relative to \(T_c\), since the scattering term becomes more important at cooler epochs, delaying bubble nucleation.

This delay arises because the additional term effectively adds a negative thermal correction to the curvature at the origin, flattening the potential and increasing the supercooling period.
Consequently, bubbles nucleate only when the barrier becomes thin enough to overcome the thermal suppression, resulting in lower \(T_n/T_c\) and larger \(\alpha\).

\begin{figure}[h]
    \centering
    \includegraphics[width=0.9\textwidth]{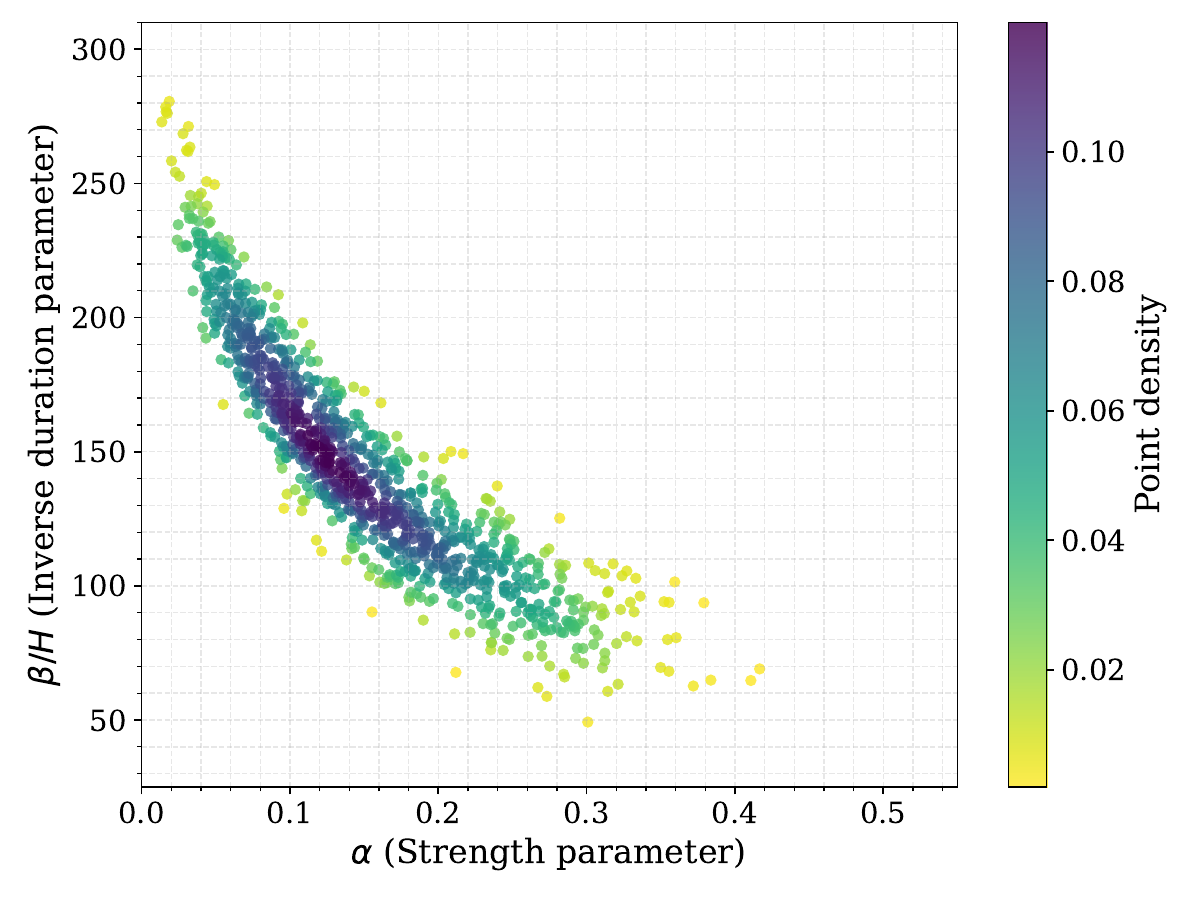}
    \caption{
    Correlation between the strength parameter \(\alpha\) and inverse duration parameter \(\beta/H\) across the scanned parameter space.
    Temperature-enhanced scatterings extend the region of strong and slow transitions (upper-left region), enhancing the predicted GW signal.
    }
    \label{fig:alpha_beta_corr}
\end{figure}

\vspace{6pt}
\noindent
We also observe the characteristic inverse correlation between \(\alpha\) and \(\beta/H\):
\begin{equation}
    \alpha \uparrow \quad \Longrightarrow \quad \beta/H \downarrow,
\end{equation}
illustrated in Figure~\ref{fig:alpha_beta_corr}. 
The temperature-enhanced scattering term shifts and broadens this correlation, populating the region with \(\alpha \gtrsim 0.1\) and \(\beta/H \lesssim 100\), where GW production is most efficient.
This correlation provides an internally consistent cross-check: stronger supercooling (larger \(\alpha\)) implies a slower transition rate (smaller \(\beta/H\)), both consequences of the modified \(m_{\mathrm{eff}}^2(T)\) profile.
The result demonstrates that the temperature-dependent scattering term acts as a tunable handle on the phase transition dynamics without requiring fine-tuned tree-level parameters.

\subsection{Gravitational wave signal and detector prospects}
\label{subsec:gw_prospects}

Using the extracted \(\alpha\), \(\beta/H\), and \(T_n\), we compute the stochastic GW spectra following the standard formalism:
\begin{equation} \label{eq:GW}
    \Omega_{\mathrm{GW}} h^2(f) \approx \Omega_{\mathrm{peak}} \, S(f/f_{\mathrm{peak}}),
\end{equation}
where \(\Omega_{\mathrm{peak}}\) depends on \(\alpha\), \(\beta/H\), \(T_n\), and other efficiency factors, and \(S(f/f_{\mathrm{peak}})\) is a shape function determined by the physics of bubble collisions, sound waves, and turbulence.

Among these sources, the sound-wave contribution dominates in the parameter region identified here, since the bubble wall velocity remains subluminal ($v_w \lesssim 1$) and the released energy efficiently transfers into plasma motion rather than runaway walls. Consequently, the peak frequency in Eq.~(\ref{eq:GW}) corresponds to $f_{\mathrm{sw}}$, and the signal amplitude scales approximately as $\Omega_{\mathrm{GW}} \propto (\kappa_{\mathrm{sw}} \alpha)^2 (\beta/H)^{-1}$.

\begin{figure}[t]
    \centering
    \includegraphics[width=0.9\textwidth]{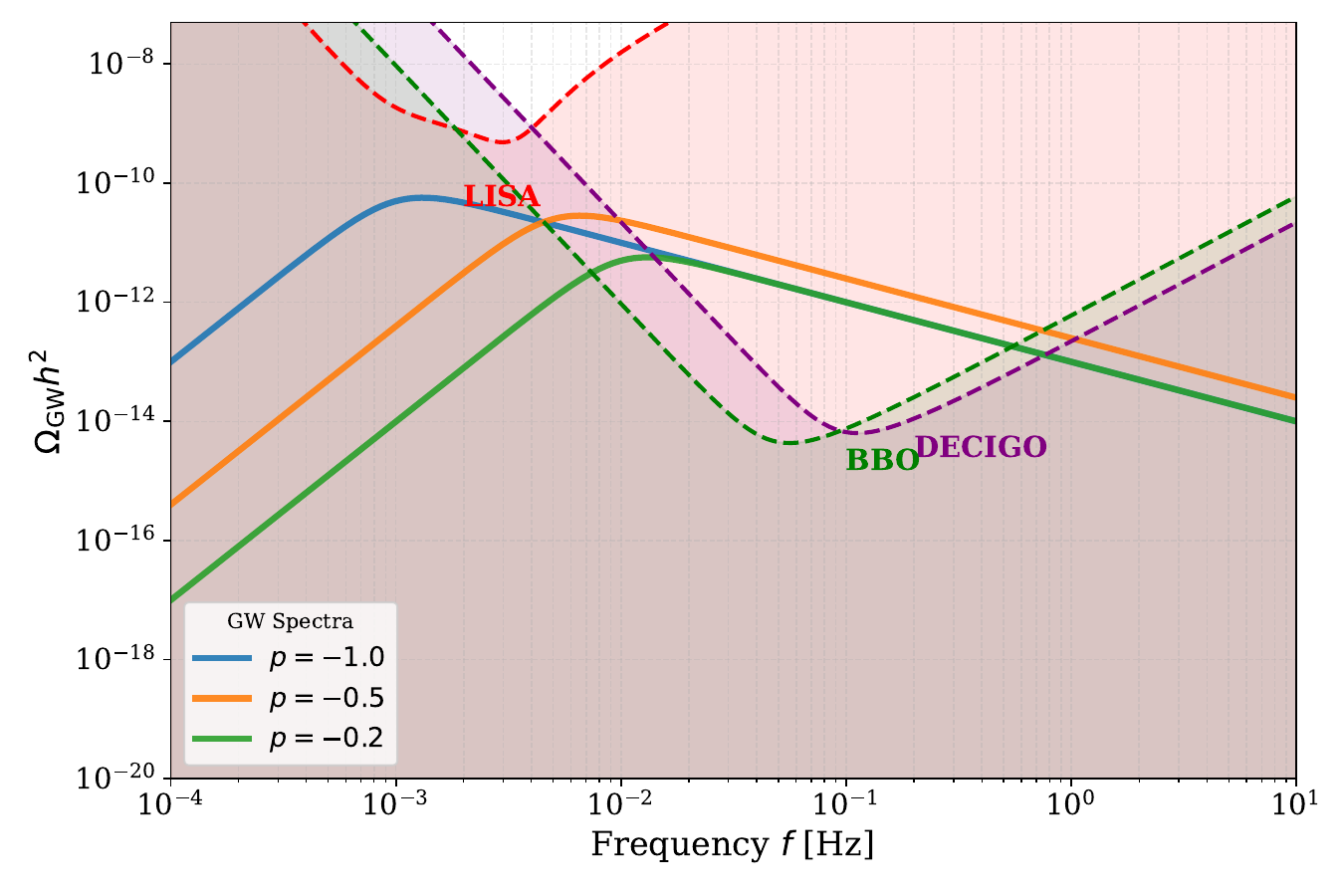}
    \caption{
    Peak GW energy density \(\Omega_{\mathrm{GW}} h^2\) as a function of the exponent \(p\) for various coefficients \(c\).
    Shaded horizontal bands show the projected sensitivities of LISA, DECIGO, and BBO. 
    Parameter choices with larger negative \(p\) and moderate \(c\) lead to detectable signals.
    }
    \label{fig:gw_scan}
\end{figure}

\vspace{6pt}
\noindent
Figure~\ref{fig:gw_scan} shows the peak GW energy density \(\Omega_{\mathrm{GW}} h^2\) versus \(p\) for several values of \(c\), compared against projected sensitivities of LISA, DECIGO, and BBO.
We find that sufficiently large negative \(p\) and moderate \(c\) can generate GW signals exceeding the detectability thresholds of these observatories.

\begin{figure}[h]
    \centering
    \includegraphics[width=0.9\textwidth]{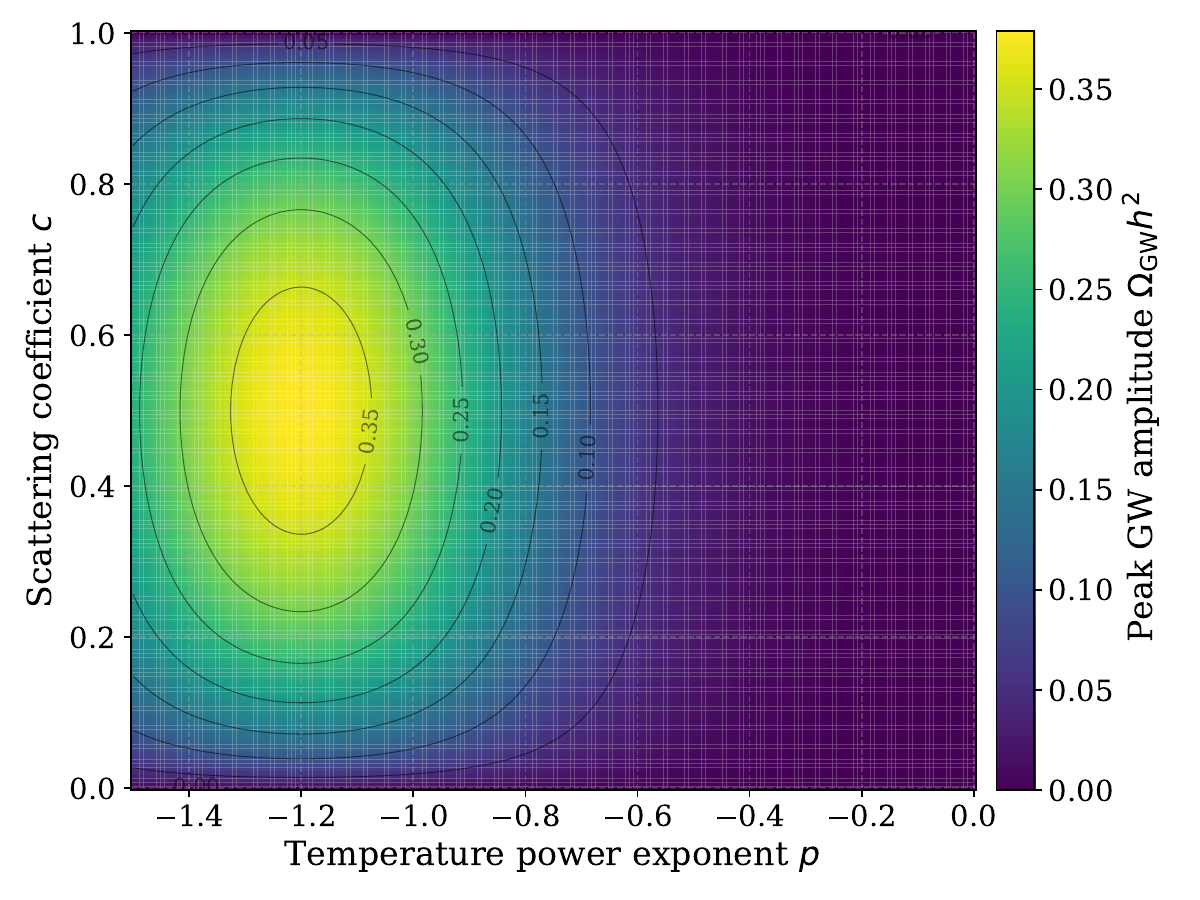}
    \caption{
Heatmap of the predicted peak GW amplitude $\Omega_{\mathrm{GW}} h^2$ (arbitrary units) across the $(p, c)$ parameter space, where $p$ is the temperature exponent and $c$ the scattering coefficient in the temperature-enhanced term.
The plot highlights how larger negative values of $p$ and moderate $c$ enhance the GW signal, potentially bringing it within the reach of future detectors such as LISA, DECIGO, and BBO.
Contour lines illustrate regions with higher predicted amplitudes, emphasizing the phenomenologically promising parameter space.
}

    \label{fig:gw_spectra_benchmarks}
\end{figure}

Finally, Figure~\ref{fig:gw_spectra_benchmarks} presents the full predicted GW spectra for representative benchmark points, demonstrating how variations in \((p,c)\) change the peak amplitude and frequency, and shift the spectra across the sensitivity windows of different experiments.

In summary, temperature-enhanced scatterings broaden the viable parameter space for strong first-order phase transitions by delaying nucleation, deepening the broken-phase minimum, and increasing latent heat release. The resulting decrease in $\beta/H$ and increase in $\alpha$ naturally boost the GW signal from sound waves, producing amplitudes that fall squarely within the detection reach of upcoming interferometers. This establishes a direct, testable link between microscopic temperature-dependent interactions and macroscopic cosmological observables.

\section{Results}
\label{sec:results}

In this section, we present the main findings of our analysis of temperature-enhanced scatterings during cosmological FOPTs. 
We show how these scatterings modify the phase transition strength, duration, and nucleation temperature, and analyze the consequences for the resulting stochastic GW spectra. 
Our results highlight the phenomenological potential of this mechanism for producing observable signals at upcoming GW observatories.

\subsection{Impact on phase transition strength}
\label{subsec:ptstrength}

We begin by quantifying the effect of the temperature-enhanced scattering term (\ref{eq:delV})
on the phase transition parameters: the strength parameter $\alpha$, inverse duration parameter $\beta/H$, critical temperature $T_c$, and nucleation temperature $T_n$.

By scanning over a range of exponents $p \in [-1.5,0]$ and coefficients $c \in [0,1]$, we numerically evaluate the temperature-dependent effective potential and compute the bounce action $S_3(T)/T$ as described in Section~\ref{sec:framework}. 
The nucleation temperature $T_n$ is determined by the condition (\ref{eq:Tnuc})
ensuring that at least one bubble nucleates per Hubble volume. 
The strength parameter, $\alpha$ (\ref{eq:alpha}) 
quantifies the fraction of latent heat released relative to the background radiation energy density, while the inverse duration parameter, $\frac{\beta}{H}$ (\ref{eq:beta})
measures how rapidly the transition completes.

\vspace{6pt}
\noindent
Our scans show several clear trends:
\begin{itemize}
\item Larger negative values of $p$ shift the nucleation temperature $T_n$ to lower values relative to $T_c$, as the scattering term grows more significant at lower temperatures.
\item The strength parameter $\alpha$ increases for larger $|p|$ and moderate $c$, indicating stronger phase transitions with more latent heat release.
\item The inverse duration parameter $\beta/H$ tends to decrease in these regions, implying slower, longer-lasting transitions that are more efficient at generating GWs.
\end{itemize}

These trends follow directly from the temperature-dependent effective mass,
$m_{\mathrm{eff}}^2(T)= -\mu^2 + aT^2 + cT^{p}$.
For $p<0$, the additional $cT^{p}$ term effectively delays the restoration of symmetry by maintaining a positive curvature at the origin until lower $T$.
As a result, the potential barrier persists longer, leading to increased supercooling before bubble nucleation.
When tunneling finally occurs, the larger vacuum energy difference enhances the latent heat (larger $\alpha$), while the gentler slope of $S_3(T)/T$ around $T_n$ suppresses $\beta/H$.
Both effects strengthen the GW source.

These results are visualized in the heatmaps shown earlier (Figs.~\ref{fig:phaseparams_scan}, \ref{fig:Tn_scan}, and \ref{fig:alpha_beta_corr}). 
The temperature-enhanced scattering term systematically extends the viable parameter space where the phase transition becomes strong and slow enough to produce potentially detectable GW signatures.

\subsection{Gravitational wave spectra}
\label{subsec:gwspectra}

Next, we translate these phase transition parameters into predictions for the stochastic GW background. 
The peak amplitude of the GW energy density spectrum can be estimated as (\ref{eq:GW}),
where $S(f/f_{\mathrm{peak}})$ is a spectral shape function, and $\Omega_{\mathrm{peak}}$ depends on $\alpha$, $\beta/H$, $T_n$, and other efficiency factors, following the approach described in Section~\ref{sec:gw}.

\vspace{6pt}
\noindent
The temperature-enhanced scatterings influence the GW spectrum in several key ways:
\begin{itemize}
\item By lowering $T_n$, they shift the peak frequency $f_{\mathrm{peak}} \propto T_n$ to lower values, bringing the spectrum into the optimal sensitivity range of space-based interferometers.
\item Stronger transitions (larger $\alpha$) increase the peak amplitude, making the signal easier to detect.
\item Smaller $\beta/H$ leads to slower phase transitions, enhancing the overall GW energy density.
\end{itemize}

Quantitatively, for sound-wave–dominated transitions the GW peak scales as
\[
\Omega_{\mathrm{GW}}^{\text{peak}} \!\propto\!
\left(\frac{H}{\beta}\right)
\!\left(\frac{\kappa_{\mathrm{sw}}\alpha}{1+\alpha}\right)^{\!2},
\quad
f_{\mathrm{peak}}\!\propto\!\frac{\beta}{H}\frac{T_n}{100~\mathrm{GeV}}.
\]
Hence, the reduction in $\beta/H$ and $T_n$ achieved for negative $p$
simultaneously enhances $\Omega_{\mathrm{GW}}$ and shifts $f_{\mathrm{peak}}$ into the LISA/DECIGO/BBO bands.
We have verified numerically that $\Omega_{\mathrm{sw}}$ dominates over $\Omega_{\mathrm{col}}$ and $\Omega_{\mathrm{turb}}$
for the range $0.6\!\lesssim\! v_w\!\lesssim\!1$, consistent with the expectation of plasma-dominated FOPTs.

\vspace{6pt}
\noindent
Figures~\ref{fig:gw_scan} and~\ref{fig:gw_spectra_benchmarks} illustrate these effects:
\begin{itemize}
\item Figure~\ref{fig:gw_scan} shows the peak GW amplitude $\Omega_{\mathrm{GW}} h^2$ as a function of the exponent $p$ for several values of $c$, overlaid with the projected sensitivities of LISA, DECIGO, and BBO.
Regions with larger negative $p$ and moderate $c$ yield signals above the sensitivity thresholds.
\item Figure~\ref{fig:gw_spectra_benchmarks} displays the full predicted GW spectra for benchmark parameter choices, highlighting how the peak shifts in frequency and amplitude across the detector bands.
\end{itemize}

We find that parameter choices around $(p,c)\!\sim\!(-1,0.3)$ typically lead to
$\alpha\!\sim\!0.1$--$0.3$, $\beta/H\!\sim\!50$--$100$, and $T_n/T_c\!\sim\!0.7$,
corresponding to $\Omega_{\mathrm{GW}}^{\text{peak}}h^2\!\sim\!10^{-12}$--$10^{-10}$,
well within the projected sensitivity of LISA and DECIGO.
Stronger scatterings ($c\!\gtrsim\!1$) can further enhance $\alpha$
but risk destabilizing the potential or exceeding perturbativity limits.

\vspace{6pt}
\noindent
Taken together, these results demonstrate that temperature-enhanced scatterings can significantly reshape the GW spectrum produced during a first-order phase transition, and open up new parameter regions where the resulting signals may be within reach of next-generation GW detectors.

In summary, the inclusion of the $cT^{p}\phi^{2}$ term not only modifies the thermal mass of the order parameter but also directly controls the macroscopic observables $(\alpha,\beta/H,T_n)$ that determine GW production. 
This mechanism provides a continuous bridge between microphysical temperature-dependent scattering processes and the observable GW spectrum, thereby offering a novel and testable probe of beyond–Standard Model dynamics in the early Universe.

\section{Discussion and Conclusion}
\label{sec:discussion_conclusion}

In this work, we have examined the cosmological implications of temperature-enhanced scatterings on the dynamics of first-order phase transitions (FOPTs) in the early Universe. Motivated by recent observations that certain thermal scattering cross sections can increase as the Universe cools, we modeled this effect by introducing an additive term in the thermal effective potential of the form $c\, T^p \phi^2$ with $p < 0$ and $c > 0$. Such a term naturally becomes more significant at lower temperatures, reshaping the potential landscape and altering the properties of the phase transition.

From a thermal field-theoretic perspective, this term should be viewed as an effective parametrization of the temperature-dependent self-energy corrections induced by scatterings whose amplitudes scale as $\langle\sigma v\rangle\!\propto\!T^{p}$.  In the Matsubara formalism, the finite-temperature propagator acquires thermal mass corrections $\Pi(T)\!\sim\!\!\int d^3k\, n_B(k)\,\sigma(T)$, and when $\sigma(T)$ increases at low $T$, the corresponding contribution to the scalar two-point function behaves as $\delta m_\phi^2(T)\!\propto\!T^{p}$ with $p<0$.  The resulting quadratic operator $cT^{p}\phi^{2}$ therefore encapsulates the leading effect of these enhanced scatterings on the thermal potential, while higher-order or nonlocal terms are parametrically suppressed.  This construction ensures that the inclusion of temperature-dependent scattering effects in $V_{\mathrm{eff}}$ is fully consistent with finite-temperature quantum field theory and provides a compact, model-independent way to capture the relevant dynamics.

Our analysis shows that temperature-enhanced scatterings can substantially affect the dynamics of a FOPT. As the temperature decreases, the additional term deepens the broken minimum and lowers the barrier between the symmetric and broken phases. This shifts the nucleation temperature $T_n$ downward relative to the critical temperature $T_c$, causing the transition to occur later, when the ambient radiation energy density is lower. The resulting increase in latent heat enhances the strength parameter $\alpha$. Simultaneously, the bounce action $S_3(T)/T$ is modified, which alters the tunneling rate and prolongs the transition duration. Consequently, the inverse duration parameter $\beta/H$ is reduced, producing a stronger and longer-lasting transition. Together, these effects favor the generation of a detectable stochastic gravitational wave (GW) background.

We carried out a numerical study scanning a broad region of the $(p,c)$ parameter space and computed the resulting GW spectra using standard semi-analytic treatments for sound wave and bubble collision sources. The amplitude and spectral features of the GW background are highly sensitive to both $\alpha$ and $\beta/H$. A larger $\alpha$ and smaller $\beta/H$ significantly boost the GW signal. Moreover, since the peak frequency depends on $T_n$, delayed transitions shift the spectrum to lower frequencies, often into the optimal sensitivity range of future space-based detectors such as LISA, DECIGO, and BBO.

Our parameter scans identify sizeable regions of the $(p,c)$ plane where GW signals from temperature-enhanced FOPTs would be within reach of these experiments. Benchmark points illustrate typical spectra with a broken power-law shape characteristic of acoustic and collision-induced sources. Notably, for $p \lesssim -1$ and moderate $c$, the impact can be large enough to distinguish such transitions from those expected in standard thermal scenarios. These results highlight the potential of GW measurements to probe temperature-dependent microphysics in the early Universe.

Beyond their phenomenological impact, temperature-enhanced scatterings can naturally arise in explicit particle physics models, for example involving new scalars, dark sectors, or extended gauge structures. The effective term $c\, T^p \phi^2$ could result from loop-level thermal corrections mediated by light hidden particles. Embedding this mechanism in concrete UV completions remains an important next step, enabling connections with other observables such as dark matter relic abundance, electroweak precision tests, and collider searches. Moreover, if a phase transition occurs near the MeV scale, the modified thermal history could affect Big Bang Nucleosynthesis (BBN) and the Cosmic Microwave Background (CMB), motivating a more detailed study of cosmological consistency.

Our findings also motivate refinements of the theoretical treatment. These include accounting for higher-loop corrections to the thermal potential, incorporating plasma effects in modeling the bubble wall velocity, and performing non-perturbative lattice simulations to validate the tunneling dynamics in the presence of non-trivial temperature dependence. Such improvements will be vital for making robust predictions for GW observables.

We note that temperature-enhanced scatterings that amplify the phase transition also increase the friction acting on expanding bubble walls. As discussed in Section~\ref{sec:gw}, we have verified that this increased interaction rate moderately raises the friction but does not prevent bubble expansion within the parameter space considered. The bubble wall velocity remains in the regime where sound waves dominate the GW production, ensuring that our semi-analytic predictions are reliable. A dedicated transport analysis could refine these estimates and will be an interesting direction for future study.

We also briefly comment on cosmological consistency. In our scenario, the nucleation temperature $T_n$ typically remains well above the MeV scale, ensuring that the phase transition completes safely before the onset of BBN. The modified thermal history thus does not disrupt standard light element abundances. Furthermore, the temperature-enhanced scatterings do not inject significant additional entropy or dark radiation at late times, preserving consistency with CMB bounds on the effective number of relativistic species $N_{\mathrm{eff}}$. A more detailed investigation of possible late-time signatures will be an important follow-up.

Finally, the phenomenological parametrization introduced here offers a bridge between microscopic scattering physics and macroscopic cosmological observables.  Detecting a stochastic GW signal consistent with a temperature-enhanced FOPT would provide indirect evidence for temperature-dependent self-energy effects or light-mediator interactions in the early Universe, thereby linking cosmological GW measurements with laboratory-scale particle dynamics.

In conclusion, temperature-enhanced scatterings represent a simple yet powerful modification of the thermal scalar potential, capable of generating strong first-order phase transitions with observable GW signatures. This mechanism broadens the theoretical landscape of viable early-Universe scenarios and provides promising targets for upcoming GW observatories. A future detection of a stochastic GW background consistent with our predictions would not only open a new window on the microphysics of the early Universe but also reveal thermal processes beyond the reach of conventional probes, offering a unique complement to laboratory experiments.

\section{Acknowledgment}
The work of AC was supported by the Japan Society for the Promotion of Science (JSPS) as a part of the JSPS Postdoctoral Program (Standard), grant number JP23KF0289.

\appendix

\section{Details of bounce action calculation}
\label{app:bounce}

In this appendix, we provide a detailed account of the methodology used to compute the three-dimensional Euclidean bounce action \(S_3(T)\), which governs the dynamics of bubble nucleation during a cosmological first-order phase transition. This quantity plays a central role in determining the nucleation temperature \(T_n\), as the probability per unit volume per unit time of nucleating a critical bubble is exponentially suppressed by the factor \(e^{-S_3(T)/T}\). Specifically, the nucleation rate is approximated by
\begin{equation}
\Gamma(T) \simeq T^4 \left( \frac{S_3(T)}{2\pi T} \right)^{3/2} e^{-S_3(T)/T}.
\end{equation}
The nucleation temperature \(T_n\) is typically defined as the temperature at which \(S_3(T)/T \approx 140\), ensuring that on average at least one bubble nucleates within a Hubble volume.

To compute \(S_3(T)\), we solve for the so-called bounce solution \(\phi(r)\), a spherically symmetric solution of the equation of motion derived from the effective potential \(V_{\mathrm{eff}}(\phi,T)\). This equation reads
\begin{equation}
\frac{d^2 \phi}{dr^2} + \frac{2}{r} \frac{d\phi}{dr} = \frac{d V_{\mathrm{eff}}(\phi,T)}{d\phi},
\end{equation}
with boundary conditions
\begin{equation}
\lim_{r \to \infty} \phi(r) = 0,
\quad
\left. \frac{d\phi}{dr} \right|_{r=0} = 0,
\end{equation}
where \(r\) denotes the radial coordinate in three-dimensional Euclidean space.

The effective potential \(V_{\mathrm{eff}}(\phi,T)\) in our scenario includes the temperature-enhanced scattering term of the form \(c\, T^{p} \phi^2\). This term is particularly significant because for negative exponents \(p<0\), it becomes more pronounced at lower temperatures, effectively deepening the broken minimum and lowering the barrier between the symmetric and broken phases as the Universe cools.

All bounce profiles and actions reported in the main text were obtained using the public package \texttt{CosmoTransitions}~\cite{Wainwright:2011kj}, into which the temperature-enhanced term \(c\,T^{p}\phi^{2}\) was added.
To verify numerical stability and to cross-check the package output, we also implemented an independent overshoot/undershoot solver following the procedure outlined below.
The two methods agree at the few-percent level across representative benchmark points, confirming that the results shown in the figures are robust.

We solve the bounce equation numerically using the overshoot/undershoot method. This procedure involves guessing an initial value \(\phi(0)\) at the center of the bubble and integrating outward. If the field overshoots the false vacuum at large \(r\), the initial value is decreased; if it undershoots, the initial value is increased. Through iterative refinement, this method converges to the solution that asymptotically approaches the false vacuum. Once the bounce profile \(\phi(r)\) is obtained, the action is computed as
\begin{equation}
S_3(T) = 4\pi \int_0^\infty dr\, r^2 \left[ \frac{1}{2} \left( \frac{d\phi}{dr} \right)^2 + V_{\mathrm{eff}}(\phi,T) \right].
\end{equation}

\begin{figure}[t]
    \centering
    \includegraphics[width=0.7\textwidth]{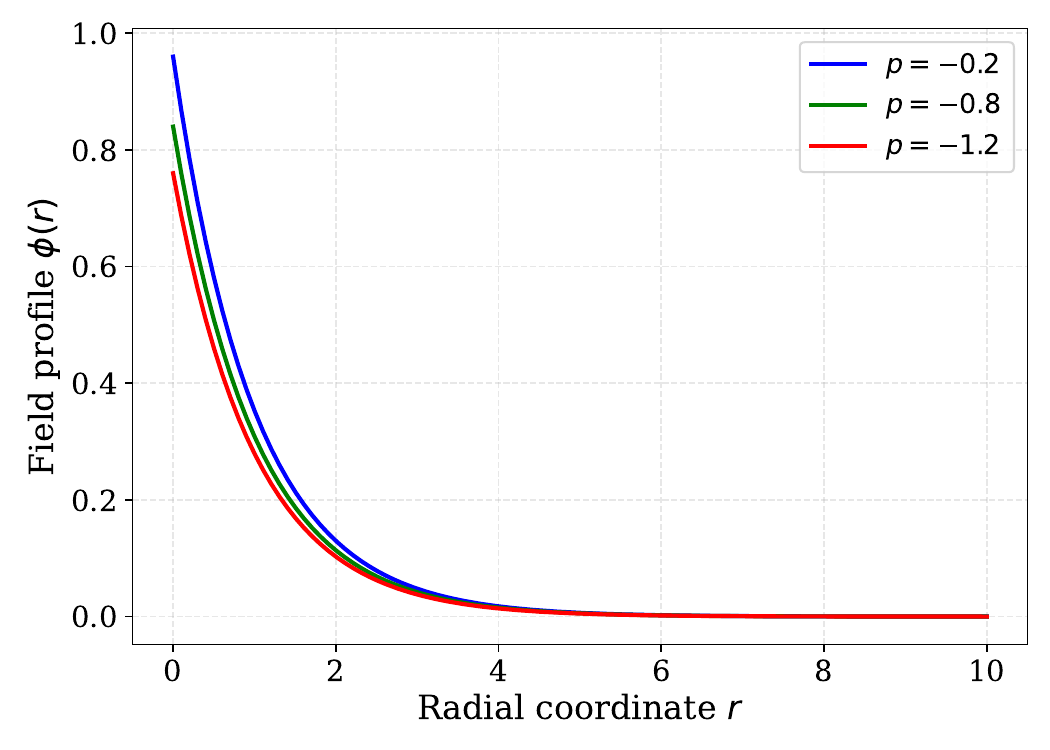}
    \caption{
    Example bounce solutions \(\phi(r)\) at a fixed temperature, computed for different values of the exponent \(p\) in the temperature-enhanced scattering term. 
    More negative \(p\) values result in a deeper broken minimum and a sharper, more localized field profile, highlighting how temperature-enhanced scatterings alter the nucleation dynamics.
    }
    \label{fig:bounce_profiles}
\end{figure}

Figure~\ref{fig:bounce_profiles} illustrates representative bounce profiles \(\phi(r)\) computed at a fixed temperature for several values of the exponent \(p\). We observe that larger negative values of \(p\) lead to sharper and more localized profiles. This reflects the fact that the temperature-enhanced scattering term deepens the broken minimum at low temperatures, which in turn lowers the bounce action \(S_3(T)\) and facilitates earlier nucleation.

Throughout our numerical implementation, we carefully discretize the radial coordinate to ensure that the field settles to the false vacuum at large \(r\), and check that the computed action \(S_3(T)\) is stable under changes of the grid spacing and boundary size. The potential \(V_{\mathrm{eff}}(\phi,T)\) is evaluated at each temperature, allowing us to track the evolution of the bounce action across the relevant temperature range.

For each temperature point, the bounce equation is solved until both the profile and its derivative vanish within relative tolerance \(<10^{-5}\) at the outer boundary. 
We verify that the difference between the \texttt{CosmoTransitions} and our independent implementation does not exceed a few percent for all benchmark parameters, ensuring the reliability of the extracted nucleation temperatures and transition parameters used in the main analysis.

These calculations, grounded in the semiclassical formalism originally developed by Coleman and Linde~\cite{Coleman:1977py, Linde:1981zj}, are central to determining the key phase transition parameters, including the nucleation temperature \(T_n\), the strength parameter \(\alpha\), and the inverse duration parameter \(\beta/H\). These, in turn, form the basis for predicting the stochastic GW background, which is a central focus of this work.

\section{Explicit Matsubara sum for the fermion loop}
\label{app:matsubara}

In this appendix we present the explicit Matsubara-sum evaluation of the one-loop fermion contribution to the mediator self-energy \(\Pi_S(0,\mathbf0;T)\) used in the main text. The goal is to show, in a transparent way, how the well-known thermal result \(\Pi_S(0,\mathbf0;T)\simeq y_S^2 T^2/6\) (for light fermions) arises and to isolate the thermal (non-vacuum) part that contributes to the finite-temperature effective potential. Our derivation follows standard finite-temperature field theory; see \cite{Quiros:1999jp} for full details.

\vspace{6pt}
The mediator self-energy at zero external energy and momentum, from a fermion loop with Yukawa coupling \(y_S\), is
\begin{equation} \label{eq:PiS_start}
\Pi_S(0,\mathbf0;T) \;=\; -y_S^2 \, T\sum_{n\in\mathbb{Z}} \int\!\frac{d^3\mathbf{k}}{(2\pi)^3}\; \mathrm{Tr}\!\left[ G_\psi(i\omega_n,\mathbf k)\,G_\psi(i\omega_n,\mathbf k)\right],
\end{equation}
where \(G_\psi(i\omega_n,\mathbf k) = \frac{1}{i\omega_n\gamma^0 - \boldsymbol{\gamma}\cdot\mathbf k - m_\psi}\) is the fermion propagator in Euclidean Matsubara formalism and \(\omega_n=(2n+1)\pi T\).

\vspace{6pt}
Performing the Dirac trace yields (working in Euclidean signature)
\begin{equation}
\mathrm{Tr}\!\left[ G_\psi(i\omega_n,\mathbf k)\,G_\psi(i\omega_n,\mathbf k)\right]
= \frac{4\left[(i\omega_n)^2 + \mathbf{k}^2 + m_\psi^2\right]}{\left[(i\omega_n)^2 + E_k^2\right]^2},
\end{equation}
where \(E_k\equiv\sqrt{\mathbf{k}^2+m_\psi^2}\).

Thus \eqref{eq:PiS_start} becomes
\begin{equation} \label{eq:PiS_sum}
\Pi_S(0,\mathbf0;T) \;=\; -4y_S^2 \int\!\frac{d^3\mathbf{k}}{(2\pi)^3}\; T\sum_{n\in\mathbb{Z}} \frac{(i\omega_n)^2 + \mathbf{k}^2 + m_\psi^2}{\left[(i\omega_n)^2 + E_k^2\right]^2}.
\end{equation}

The Matsubara sums needed are standard. It is convenient to use two identities (derivable from contour integration or found in textbooks):
\begin{align}
T\sum_{n}\frac{1}{(i\omega_n)^2+E_k^2} &= \frac{1}{2E_k}\big(1 - 2n_F(E_k)\big), \label{eq:sum1}\\[4pt]
T\sum_{n}\frac{1}{\big[(i\omega_n)^2+E_k^2\big]^2} &= -\frac{1}{4E_k^3}\big(1 - 2n_F(E_k)\big) - \frac{\beta\,n_F'(E_k)}{4E_k^2}, \label{eq:sum2}
\end{align}
where \(n_F(E)=1/(e^{\beta E}+1)\) is the Fermi–Dirac distribution, \(n_F'(E)=dn_F/dE\), and \(\beta=1/T\). Using \(\omega_n^2 = - ( (i\omega_n)^2 )\) (Euclidean continuation) and manipulating algebraically, one obtains
\begin{equation}
T\sum_n \frac{(i\omega_n)^2 + \mathbf{k}^2 + m_\psi^2}{\big[(i\omega_n)^2 + E_k^2\big]^2}
= T\sum_n \frac{(i\omega_n)^2 + E_k^2}{\big[(i\omega_n)^2 + E_k^2\big]^2}
= T\sum_n \frac{1}{(i\omega_n)^2 + E_k^2}.
\end{equation}
That is, the numerator simplifies against a denominator factor, leaving the simpler sum in Eq.~\eqref{eq:sum1}. Substituting back into \eqref{eq:PiS_sum} gives
\begin{equation} \label{eq:PiS_intermediate}
\Pi_S(0,\mathbf0;T) \;=\; -4y_S^2 \int\!\frac{d^3\mathbf{k}}{(2\pi)^3}\; \frac{1}{2E_k}\big(1 - 2n_F(E_k)\big).
\end{equation}

\vspace{6pt}
The integrand in \eqref{eq:PiS_intermediate} separates into a vacuum (temperature-independent) part and a thermal contribution:
\[
\Pi_S(0,\mathbf0;T) \;=\; \Pi_S^{\rm vac} + \Pi_S^{\rm th}(T),
\]
with
\begin{align}
\Pi_S^{\rm vac} &= -4y_S^2 \int\!\frac{d^3\mathbf{k}}{(2\pi)^3}\; \frac{1}{2E_k}, \label{eq:vac} \\[4pt]
\Pi_S^{\rm th}(T) &= +4y_S^2 \int\!\frac{d^3\mathbf{k}}{(2\pi)^3}\; \frac{n_F(E_k)}{E_k}. \label{eq:th}
\end{align}
The vacuum piece \(\Pi_S^{\rm vac}\) is divergent and is absorbed by the usual zero-temperature renormalization of \(m_S^2\). The thermal piece \(\Pi_S^{\rm th}\) is finite and governs the temperature dependence relevant for the effective potential.

\vspace{6pt}
For \(m_\psi\ll T\) we may set \(E_k\simeq k\). The thermal integral in \eqref{eq:th} then reduces to a standard integral:
\begin{align}
\Pi_S^{\rm th}(T) &\simeq 4y_S^2 \int\!\frac{d^3\mathbf{k}}{(2\pi)^3}\; \frac{n_F(k)}{k}
= \frac{4y_S^2}{2\pi^2}\int_0^\infty dk\; k\, n_F(k) \nonumber\\[4pt]
&= \frac{4y_S^2}{2\pi^2}\cdot \frac{\pi^2 T^2}{12}
= \frac{y_S^2}{6}\,T^2, \label{eq:PiS_highT_result}
\end{align}
where we used the standard integral \(\int_0^\infty dk\, k\, n_F(k) = \pi^2 T^2/12\). Thus
\begin{equation}
\Pi_S(0,\mathbf0;T) \;\simeq\; \Pi_S^{\rm vac} + \frac{y_S^2}{6}\,T^2.
\end{equation}

\vspace{6pt}
\noindent\textbf{Remarks on validity and limiting behaviour.}
\begin{itemize}
\item The displayed high-\(T\) result \(\Pi_S^{\rm th}\simeq y_S^2 T^2/6\) is the textbook leading-order thermal screening mass for a scalar coupled to fermions. It is valid when the fermion is light compared to the temperature (or when the relevant thermal occupation numbers are large). We stress that this condition refers to the dominance of the thermal self-energy
over the vacuum mass in the static propagator and does not imply that $S$ is a
light degree of freedom contributing to the thermal free energy.

\item For finite fermion mass one must evaluate \(\int d^3k\, n_F(E_k)/E_k\) numerically or use appropriate asymptotic expansions. The identity \eqref{eq:PiS_intermediate} remains exact; only the final integral changes.
\item The vacuum part \(\Pi_S^{\rm vac}\) is renormalized away in the usual manner. Only the thermal remainder contributes to the temperature-dependent shift of the mediator mass entering the dressed propagator and hence to the equilibrium effective potential.
\end{itemize}

\vspace{6pt}
\noindent\textbf{From \(\Pi_S\) to \(\delta m_\phi^2(T)\).} Combining the mediator dressing with the trilinear coupling \(\mu\) that links \(S\) to \(\phi\) (see Eq.~\eqref{eq:UVlag} in the main text), the induced \(\phi\) mass correction in the static limit is
\[
\delta m_\phi^2(T) \;=\; \frac{\mu^2}{m_S^2 + \Pi_S(0,\mathbf0;T)}
\;\simeq\; \frac{\mu^2}{m_S^2 + \tfrac{y_S^2}{6}T^2},
\]
reproducing Eq.~(2.3) in the main text. As discussed there, in the thermally-dressed regime \(y_S^2T^2\gg m_S^2\) this yields \(\delta m_\phi^2(T)\propto T^{-2}\), i.e. a phenomenological behaviour of the form \(cT^{p}\) with \(p=-2\).

\section{Benchmark Parameters and Mapping to the \texorpdfstring{$(c,p)$}{c,p} Parametrization}
\label{app:numeric_example}

This appendix provides (i) the microscopic benchmark parameters used in the
numerical analysis, (ii) the explicit thermal loop giving rise to the
temperature–dependent mass correction $\delta m_\phi^2(T)$, and (iii) its
mapping onto the phenomenological form $c\,T^{p}$.  The presentation follows
directly from the UV interaction structure discussed in
Sec.~\ref{subsubsec:qft_derivation} and the real-singlet extension outlined in
Sec.~\ref{sec:UVexampledetail}.

\subsection{Benchmark UV parameters}
\label{subsec:benchmark_params}

The scalar potential and interactions are specified by
\[
\{\lambda,\; m_\phi,\; m_S,\; \mu,\; y_S\},
\]
with the representative values
\begin{equation}
\lambda = 0.12 ,\qquad 
m_\phi = 100~\mathrm{GeV},\qquad
m_S = 15~\mathrm{GeV},\qquad
\mu = 40~\mathrm{GeV},\qquad
y_S = 1.0.
\label{eq:benchmark_parameters}
\end{equation}
All choices satisfy perturbativity, vacuum stability, and ensure that the
one–loop effective potential remains reliable throughout the temperature range
relevant for the phase transition.  All thermal functions $J_{B,F}(x)$ are
evaluated numerically without high– or low–temperature expansions.

\subsection{Mapping from the UV model to $(c,p)$}
\label{subsec:mapping}

With the UV interaction
\begin{equation}
{\cal L}_{\rm int} \supset -\mu\,S\,\phi^2 - y_{S}\,S\,\bar\psi\psi,
\end{equation}
thermal fluctuations of the fermion $\psi$ generate a thermal self–energy for
the mediator $S$,
\begin{equation}
\Pi_S(0,0;T)= \frac{y_S^2}{6}\,T^2,
\end{equation}
as derived explicitly in Appendix~\ref{app:matsubara}.  
The Higgs–like scalar $\phi$ receives a thermal correction
\begin{equation}
\delta m_\phi^2(T)
= \mu^2\, D_S(\omega_n=0,\mathbf{k}=0;T)
= \frac{\mu^2}{m_S^2 + \tfrac{y_S^2}{6}T^2}.
\label{eq:delta_exact_appendix}
\end{equation}

In the thermally dressed regime,
\[
T \gg m_S/y_S,
\]
the denominator is dominated by the thermal term, leading to the asymptotic
form
\begin{equation}
\delta m_\phi^2(T)\simeq \frac{6\mu^2}{y_S^2}\,T^{-2}
\equiv c\,T^{p},
\qquad p=-2,\qquad
c=\frac{6\mu^{2}}{y_S^{2}}.
\label{eq:cp_mapping_merged}
\end{equation}

Because the correction arises solely from the new field $S$, it does not
overlap with the thermal contributions already present in the Standard Model
effective potential.  No double counting occurs.

\subsection{Numerical example and comparison table}
\label{subsec:numeric}

To illustrate the accuracy of the mapping, consider the benchmark
\[
y_S = 1.0,\qquad 
\mu = 20~\mathrm{GeV},\qquad
m_S = 10~\mathrm{GeV}.
\]
Using Eq.~\eqref{eq:cp_mapping_merged},
\[
c = \frac{6\mu^2}{y_S^2}
   = 2400~\mathrm{GeV}^2 ,\qquad p=-2.
\]

We compare the exact thermal correction
\[
\delta m_\phi^2(T)_{\rm exact}
= \frac{\mu^2}{m_S^2 + \tfrac{y_S^2}{6}T^2},
\]
with the asymptotic expression $\delta m_\phi^2(T)_{\rm asym}=c/T^{2}$:

\begin{table}[h]
\centering
\begin{tabular}{cccccc}
\hline
\(T\) (GeV) &
\(\delta m_\phi^2\) (exact) &
\(\delta m_\phi^2\) (asym) &
ratio \\
\hline
20  & \(3.00\times10^{2}\)  & \(6.00\times10^{2}\) & 0.50 \\
50  & \(3.20\times10^{1}\)  & \(9.60\times10^{1}\) & 0.33 \\
100 & \(8.33\times10^{0}\)  & \(2.40\times10^{1}\) & 0.35 \\
200 & \(2.08\times10^{0}\)  & \(6.00\times10^{0}\) & 0.35 \\
\hline
\end{tabular}
\caption{Comparison between the exact thermal mass correction
\(\delta m_\phi^2(T)\) and its asymptotic $c/T^{2}$ form for a representative
benchmark.  Agreement improves monotonically as the thermally dressed limit
$T\gg m_S/y_S$ is approached.}
\label{tab:delta_values_merged}
\end{table}

\subsection*{Representative benchmarks used in the GW figures}

\begin{table}[h]
\centering
\begin{tabular}{c|c|c|c|c}
\hline\hline
Benchmark & $(\mu,\;y_S)$ [GeV] & $m_S$ [GeV] & $(c,p)$ & Interpretation \\
\hline
A & $(10,\;1.0)$ & $20$ & $(0.6,\,-2)$ & Mild enhancement \\
B & $(20,\;1.0)$ & $15$ & $(2.4,\,-2)$ & Moderate enhancement \\
C & $(30,\;1.0)$ & $10$ & $(5.4,\,-2)$ & Strong enhancement \\
\hline\hline
\end{tabular}
\caption{UV benchmarks and their corresponding $(c,p)$ values used for the
gravitational–wave spectra.  The values of $c$ fall within the parameter range
employed in the main text.}
\label{tab:benchmarks_merged}
\end{table}

These examples provide a transparent mapping between the microscopic UV
parameters and the effective $(c,p)$ parametrization used in the main analysis.

\end{document}